\documentclass{article} 
\usepackage{iclr2021_conference,times}
\iclrfinalcopy

\usepackage{amsmath,amsfonts,bm}

\def\eqref#1{equation~\ref{#1}}

\def\ceil#1{\lceil #1 \rceil}

\def\1{\bm{1}}

\DeclareMathAlphabet{\mathsfit}{\encodingdefault}{\sfdefault}{m}{sl}
\SetMathAlphabet{\mathsfit}{bold}{\encodingdefault}{\sfdefault}{bx}{n}

\usepackage{hyperref}
\usepackage{url}

\usepackage{booktabs} 
\usepackage{graphicx}
\usepackage{subfig}
\usepackage{amsmath}
\usepackage{tikz}
\usepackage{pgfplots}
\usepackage{makecell}
\pgfplotsset{compat=newest}
\usepgfplotslibrary{groupplots}
\usepgfplotslibrary{dateplot}

\usepackage{xspace}
\newcommand{\wtf}{\textsc{not-so-big-GAN}\xspace}

\newcommand{\wt}{\textsc{nsb-GAN}\xspace}
\newcommand{\wtw}{\textsc{nsb-GAN-W}\xspace}
\newcommand{\wtp}{\textsc{nsb-GAN-P}\xspace}

\title{\wtf: Generating High-Fidelity Images on Small Compute with Wavelet-based Super-Resolution}

\author{%
Seungwook Han\thanks{equal contribution} \\
  MIT-IBM Watson AI Lab\\
  IBM Research\\
  \texttt{seungwook.han@ibm.com} \\
   \And
   Akash Srivastava* \\
   MIT-IBM Watson AI Lab\\
   IBM Research\\
  \texttt{akash.srivastava@ibm.com} \\
   \AND
   Cole Hurwitz* \\
   MIT-IBM Watson AI Lab \\
   University of Edinburgh \\
   \texttt{cole.hurwitz@ed.ac.uk} \\
   \And
   Prasanna Sattigeri \\
   IBM Research \\
   \texttt{psattig@us.ibm.com} \\
   \And
   David D. Cox \\
   MIT-IBM Watson AI Lab \\
   IBM Research \\
   \texttt{david.d.cox@ibm.com} \\
   }

\begin{document}

\maketitle

\begin{abstract}

State-of-the-art models for high-resolution image generation, such as BigGAN and VQVAE-2, require an incredible amount of compute resources and/or time (512 TPU-v3 cores) to train, putting them out of reach for the larger research community. On the other hand, GAN-based image super-resolution models, such as ESRGAN, can not only upscale images to high dimensions, but also are efficient to train. In this paper, we present \wtf (\wt), a  simple yet cost-effective two-step training framework for deep generative models (DGMs) of high-dimensional natural images. First, we generate images in low-frequency bands by training a \textit{sampler} in the wavelet domain. Then, we super-resolve these images from the wavelet domain back to the pixel-space with our novel wavelet super-resolution \textit{decoder} network. Wavelet-based down-sampling method preserves more structural information than pixel-based methods, leading to significantly better generative quality of the low-resolution sampler (e.g., 64$\times$64).
Since the sampler and decoder can be trained in parallel and operate on much lower dimensional spaces than end-to-end models, the training cost is substantially reduced. On ImageNet 512$\times$512, our model achieves a Fréchet Inception Distance (FID) of 10.59 -- beating the baseline BigGAN model -- at half the compute (256 TPU-v3 cores).

\end{abstract}

\section{Introduction}
Generative modeling of natural images has achieved great success in recent years \citep{vae,gan,wgan,spn,sagan}. Advancements in scalable computing and theoretical understanding of generative models \citep{spectral, sagan, wgangp, lars1,lars2,kevin,fgan,veegan,gram, stylegan}, have, for the first time, enabled the state-of-the-art techniques to generate photo-realistic images in higher dimensions than ever before \citep{biggan,vqvae2,stylegan}. Yet, generating high-dimensional complex data, such as ImageNet, still remains challenging and extremely resource intensive.
At the forefront of high-resolution image generation is BigGAN \citep{biggan}, a generative adversarial network (GAN) \citep{gan} that tackles the curse of dimensionality (CoD) head-on, using the latest in scalable GPU-computing. This allows for training BigGAN with large mini-batch sizes (e.g., 2048), which greatly helps to model highly diverse, large-scale datasets like ImageNet. But, BigGAN's ability to scale to high-dimensional data comes at the cost of a hefty compute budget. A standard BigGAN model at 256$\times$256 resolution can require up to a month or more of training time on as many as eight Tesla V100 graphics processing units (GPUs). This compute requirement raises the barrier to entry for using and improving upon these technologies as the wider research community may not have access to any specialized hardware (e.g., Tensor processing units (TPUs) \citep{tpu}. The environmental impact of training large-scale models can also be substantial as training BigGAN on 512$\times$512 images with 512 TPU cores for two days reportedly used as much electricity as the average American household does in about six months \citep{env}.

\begin{figure}[h]
    \centering
    \includegraphics[width=\textwidth]{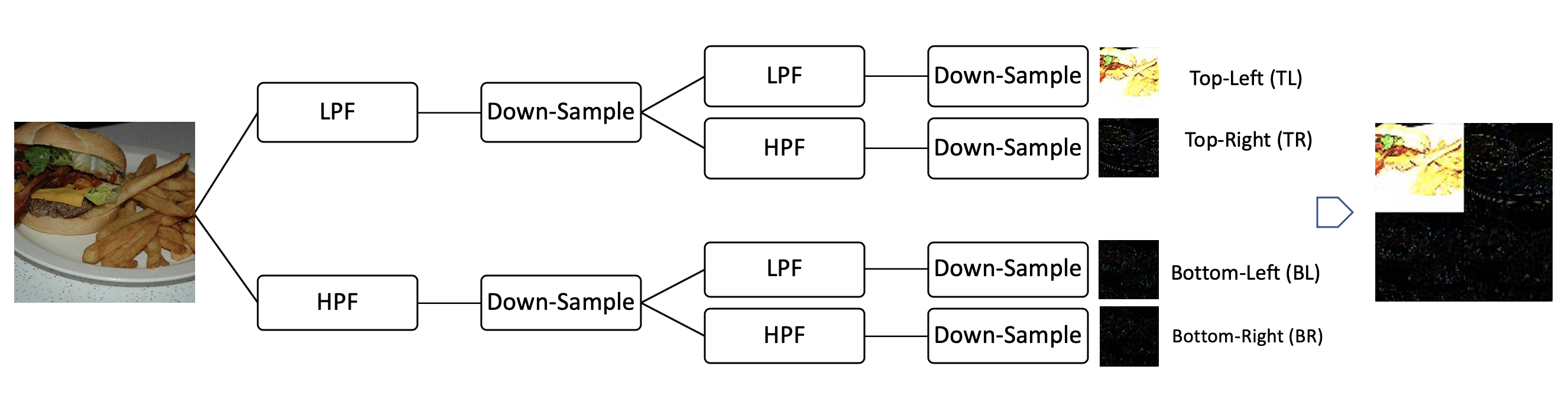}
    \caption{Wavelet transformation consists of a low-pass and a high-pass filter, followed by a down-sampling step that splits the image into two equal-sized patches. Each of these two patches undergo the same operation again resulting in four equal-sized patches, TL, TR, BL and BR. }
    \label{fig:wt}
\end{figure}

Motivated by these problems, we present \wtf (\wt), a small compute training alternative to BigGAN, for class-conditional modeling of high-resolution images. In end-to-end generative models of high-dimensional data, such as VQVAE-2 \citep{vqvae2} and \cite{progressive}, the lower layers transform noise into low resolution images, which are subsequently upscaled i.e. \textit{super-resolved} to higher dimensions in the higher layers. Based on this insight, in \wt we propose to split the end-to-end generative model into two separate neural networks, a sampler and an up-scaling decoder that can be trained in parallel on much smaller dimensional spaces. In turn, we drastically reduce the compute budget of training. This split allows the sampler to be trained in up to 16-times lower dimensional space, not only making it compute efficient, but also alleviating the training instability of end-to-end approaches. To this end, we propose wavelet-space training of GANs. As compared to pixel-based interpolation methods for down-sampling images, wavelet-transform (WT) \citep{wt0,wt,wt1} based down-sampling preserves much more structural information, leading to much better samplers in fairly low resolutions \citep{6922406}. When applied to a 2D image, wavelet transform slices the image into four equally-sized image-like patches along different frequency bands. This process can be recursively applied multiple times in order to slice a large image into multiple smaller images, each representing the entire image in different bandwidths. This is diagrammatically shown in Figure \ref{fig:wt}. Here, the top-left patch (TL) lies in the lowest frequency band and contains most of the structure of the original image and therefore the only patch preserved during downsampling. The highly sparse top-right (TR), bottom-left (BL) and bottom-right (BL) patches lie in higher bands of frequency and are therefore dropped. But wavelet-space sampling prohibits the use of existing pixel-space super-resolution models, such as \cite{srgan,esrgan}, to upscale the samples. Thus, we introduce two wavelet-space super-resolution \textit{decoder} networks that can work directly with wavelet-space image encoding, while matching the performance of equivalent pixel-space methods. Training our decoders is extremely compute efficient (e.g., 3 days on the full ImageNet dataset), and, once trained on a diverse dataset like ImageNet, can generalize beyond the original training resolution and dataset.

Our main contributions are the following: 
\begin{itemize}
\itemsep0em 
    \item We introduce a simple training framework \wt for large generative models and show that super-resolving low-resolution samplers reduces the cost of training by orders of magnitude without sacrificing image quality and even outperforms the baseline BigGAN model at higher resolution (e.g., $512 \times 512$).
    \item In addition to a simple pixel-based version of \wt, we introduce wavelet-space training of DGMs and demonstrate that it leads to better image quality compared to pixel-space \wt model on average.
    \item While conceptually simple (especially in the pixel-space), training large models to work in tandem requires careful engineering. Therefore, we make our entire \wt code-base publicly available to the wider research community.
\end{itemize}

\section{Background and Related Work}
 Given a set $X = \{x_i \vert \forall i \in \{1, \dots N\}, x_i \in \mathbb{R}^D \}$ of samples from the true data distribution $p(x)$, the task of deep generative modeling is to approximate this distribution using deep neural networks. All generative models of natural images assume that $p(x)$ is supported on a $K$-dimensional, smaller manifold of $\mathbb{R}^D$. As such, they model the data using a conditional distribution $p_\theta(x|z)$, where $\theta$ represents the model parameters and $Z \in \mathbb{R}^K$ is the latent variable with the prior $p(z)$. $Z$ is marginalized to obtain the log-likelihood (LLH) $\log p_\theta(x) = \log \int p(z) p_\theta(x|z) dz$ of $X$ under the model. Generative model are trained by either maximizing this LLH or its lower-bound.  
There are two types of deep generative models, explicit models like variational autoencoders (VAE) \citep{vae} that have an explicit form for the LLH and implicit models such as GANs \citep{gan} that do not have such a form. VAEs use a decoder network $D_\theta$ to parameterize the model distribution $p_\theta(x|z)$ and an encoder network $E_\phi$ to parameterize a variational distribution $q_\phi(z|x)$ as an approximation to the true posterior under the model. They can then be trained using the evidence lower-bound (ELBO), $\log p_\theta(x) \geq - \text{KL}[q_\phi(z|x) \Vert p(z)] + \int q_\phi(z|x) \log p_\theta(x|z)dz.$
KL here refers to the Kullback–Leibler divergence between $q_\phi(z|x)$ and $p(z)$.
Unlike VAEs, GANs do not make any assumptions about the functional form of the conditional distribution $p_\theta(x|z)$ and, therefore, do not have a tractable likelihood function. Instead, they directly attempt to minimize either a variational lowerbound to a chosen $f$-divergence \citep{fgan, veegan} or integral probability metric (IPM) \citep{wgan,mmdgan,gram} between the model distribution $p_\theta(x)$ and $p(x)$ using binary classifier based density ratio estimators. 
 Though the idea of multiscale modeling in GANs has been considered throughly in \cite{denton2015deep,progressive,stylegan,liu2019acceleration}, our method is more closely related to VQVAE-2 \citep{vqvae2} and the Subscale Pixel Network (SPN) \citep{spn} models. VQVAE-2 is a hierarchical version of the VQVAE model \citep{vqvae} that unlike VAEs, does not impose a KL-based regularization on the latent space. Instead, VQVAE/VQVAE2 models use vector quantization as a strong functional prior. As such, sampling from the latent space is not straightforward and requires a large compute budget to train auto-regressive PixelSNAIL \citep{pixelsnail} models to estimate the density on the latent space to generate samples. 
\citep{spn} shows that it is easier to generate high-resolution images by splitting the process into two steps. In step one, they slice the image into small patches and then learn to generate these patches using a patch-level autoregressive model. This step allows for capturing the general structure of the image but misses the finer details. Therefore, in step two, they train another network that learns to fill in the missing details. The SPN model has a significant shortcoming, however, as its patch-level autoregressive approach fails to capture long term dependencies (e.g. eyes on a face looking in different directions). Similar to SPN's multi-scale approach, \wt decouples structure learning from adding details. Unlike SPN, however, it does not suffer from the long-term dependency problem. This is because \wt uses wavelet transformation to slice the image, which naturally decouples the structure (TL) from the details (TR, BL, BR) in the image. We demonstrate this difference diagrammatically in the Appendix \ref{sec:sl}. More importantly, wavelet transform alleviates the need for autoregressive modeling of the patches as each patch represents the entire image in different frequency bands. 

\paragraph{Image Super Resolution} Since the pioneering work of SRCNN \citep{srcnn} to tackle the problem of single image super-resolution, recovering a high-resolution image from a low-resolution image, with a deep neural network, deep convolutional neural network approaches have demonstrated great success reconstructing realistic high-resolution images. ESRGAN \citep{esrgan} is the state-of-the-art GAN model that majorly builds on SRGAN, RDN, and EDSR \citep{srgan, rdn, edsr}. Improving upon its predecessor SRGAN, it modifies the SRGAN's architecture from SRResNet to Residual-in-Residual Dense Block (RRDB) without batch normalization, finetunes the perceptual loss implemented with the VGG-19 model \citep{vgg}, and utilizes a relativistic GAN-like adversarial loss to predict relative realness. First, only the generator is trained as a Peak Signal-to-Noise Ratio (PSNR)-oriented model with L1 loss. Then, the GAN is trained as a whole, initialized from this pre-training, with carefully balanced L1, perceptual, and adversarial losses. With such, ESRGAN is able to reconstruct realistic high-frequency texture and details missing in the low-resolution images.

\section{Method}
\wt training introduces a fairly simple change to the original BigGAN model. Instead of training the generator directly on the full dimensionality of the data, $D=256$ (say, for ImageNet dataset), we train the generator in $K=64$. Then, in order to upscale the $64\times64$ image back to $256$, we train an up-scaling decoder. We now describe two specific instances of the \wt framework, \wtw and \wtp in detail.
\subsection{\wtw}
\wtw is best described as an autoencoder that operates in the frequency domain instead of the usual pixel space. It is comprised of a deterministic encoder, a learned decoder and learned prior network, i.e. sampler.  These three components form a full generative model  that  can  produce  high-quality,  high-resolution  samples  with  a  week  of  training on commodity hardware.
\subsubsection{Deterministic Encoder.}
The \wtw encoder is a deterministic function that recursively applies wavelet transform to the input image, \textit{retaining only the TL patches at each level}. 
Each TL patch is a quarter of the size of the previous TL patch, resulting in a pyramid-like stack of 2D patches after multiple transformations. The last TL contains the lowest frequencies and, therefore, the most structural information about the image. We use it as a compressed, lossy representation of the input image in our encoding step. 

For a $N^2$-dimensional image $X$, let us denote the matrix obtained after wavelet transform as $W(X)$, which is a $N \times N$ block matrix with the following structure,
\begin{align}
    W(X)=
    \left[
    \begin{array}{c|c}
    W_{1,1}(X) & W_{1,2}(X) \\
    \hline
    W_{2,1}(X) & W_{2,2}(X)
    \end{array}
    \right]. \nonumber
\end{align}
Here, with a slight abuse of notation, $W_{1,1}(X)$ represents the \big($\frac{N}{2} \times \frac{N}{2}$\big)-dimensional TL patch of image X. \wt's encoder $\mathcal{E}$ can then be defined recursively as follows,
\begin{align}
\label{eq:encode}
    \mathcal{E}_0(X) &= X \nonumber\\
    \mathcal{E}_{l}(X) &= W_{1,1}(\mathcal{E}_{l-1}(X)), \quad 1 \leq l \leq L
\end{align}
where $l$ is the number of wavelet transforms applied to the input image. As can be seen, the size of the retained TL patch decreases exponentially with $l$. 

\subsubsection{Decoder}
After wavelet transform, the original image can be deterministically recovered from the TL, TR, BL, and BR patches using IWT. \wtw, however, discards the high-frequency patches during its encoding step, rendering a fully deterministic decoding impossible. Thus, the \wt decoder first learns to recover the missing high-frequency patches (using a neural network) and then deterministically combines them using IWT to reconstruct the original input. We now present two \wtw decoder designs based on ResNet \citep{resnet, srgan} and UNet \citep{unet} architectures respectively. 

\paragraph{ESRGAN-W}
Since \wtw sampler generates samples in the wavelet domain, we modify the basic ESRGAN architecture \citep{esrgan} to allow it to accept images in the wavelet-domain, up-scale them, and then project them back to the pixel-space. 
Specifically, we replace the generator in ESRGAN with a 16-block-deep SRResNet without batch normalization \citep{srgan}, denoted by $f_{\theta}$, and replaced the bilinear interpolation based up-scaling with IWT. We can then define the decoder as,
\begin{align}
\label{eq:decode}
     D(W_{1,1}^l;\Theta) &= \text{IWT}\Bigg(
     \left[
        \begin{array}{c|c}
        W_{1,1}^l & \pmb{0} \\
        \hline
        \pmb{0} & \pmb{0}
        \end{array}
     \right]
     \Bigg) + f_{\theta}(W_{1,1}^l).
\end{align}

This design not only allows us to directly input wavelet-encoded images in the decoder, but also lets us use the carefully balanced GAN-based training of ESRGAN since our decoder still operates in the pixel space to recover the missing higher frequencies. Similarly to ESRGAN, we pre-train SRResNet with only L1 Loss, initialize the generator at this state, and then start the full adversarial training with perceptual loss and adversarial loss. We refer to these two models as \textbf{ResNet-W} and \textbf{ESRGAN-W}, respectively. We defer further training details of the decoder to the Appendix \ref{sec:esrgan_decoders}.

\paragraph{UNet-Decoder}
While the ESRGAN-W decoder works well with wavelet-encoded inputs, it does not take full advantage of the compression that wavelet space modeling brings about. Therefore, we also introduce a UNet-based decoder that takes full advantage of the wavelet decomposition of 2D data to completely bypass the original dimensionality of the target image. Due to space limitations, we defer the details of this design to the Appendix \ref{sec:unet_decoder}.

\subsubsection{Sampler: Learned Prior Network}
The functional prior imposed by our deterministic encoder leads to a highly structured representation space made up of low frequency TL patches of images. In order to generate from \wtw, one must be able to draw samples from this space. We posit that, compared to sampling from equivalently-sized representation spaces for AE and VAEs, it is easier to sample from a low-dimensional, image-like latent space using generative methods such as GANs, as they repeatedly have been shown to excel in learning to samples from image distributions. Therefore, we train a BigGAN sampler on this representation space. As the dimensionality is considerably lower than the original image, it only takes a single week and two Tesla V100 GPUs to to train the sampler. Since the values of the TL patch may not lie in the $[0,1]$, range, care needs to be taken in order to normalize the encoded samplers properly as failing to do so prevents the BigGAN sampler to learn the correct distribution. 

\paragraph{Pre-trained Sampler.} 
It is possible to use a pre-trained BigGAN model at a lower resolution within \wtw to generate higher resolution images. Consider a pre-trained BigGAN model that can generate 128$\times$128 samples in the pixel-space. By simply projecting the samples into the frequency-space using our deterministic encoder (down to 64 $\times$ 64) and then up-scaling it through our decoder network, one can generate samples at 256$\times$256 without any significant loss in quality. In fact, these samplers can outperform the end-to-end baseline BigGAN model not only in compute requirement, but also in terms of the FID at 512$\times$512 resolution.

\subsection{\wtp}
\wtp is the pixel counterpart of \wtw. Instead of encoding, generating, and decoding in the wavelet space, it operates in the pixel space. The deterministic encoder sub-samples the original image in the pixel space to generate a low-resolution latent representation. Specifically, we realize the sub-sampling as smoothing with a Gaussian filter and sub-sampling with respect to the upscaling factor $r$ \citep{duf}. For the \wtp decoder, we use the ESRGAN model with the generator replaced with SRResNet without batch normalization, as in \wtw, which we refer to as \textbf{ESRGAN-P} and the BigGAN model on low-dimensional space for the prior network. The training procedure is same as in ESRGAN-W, where first we pre-train SRResNet with only L1 loss (namely, \textbf{ResNet-P}) and then start adversarial training. Similar to the \wtw model, we also have a UNet-version of the decoder but it reconstructs directly in the original data dimensionality like the ESRGAN based decoder. Further details are in Appendix \ref{sec:esrgan_decoders}.

\subsection{Training}
As before, let $X$ be the dataset of high-dimensional natural images. 
The \wt encoder can be defined as a deterministic function $\mathcal{E}:\mathbb{R}^D \mapsto \mathbb{R}^K$ that uses the WT to produce $Z = \{z_j \vert \forall j \in \{1, \dots N\}, z_j \in \mathbb{R}^K \}$.  Using this paired dataset $\{X,Z\}$, we treat \wt as a fully observable model and train the decoder function $D_\theta: \mathbb{R}^K \mapsto \mathbb{R}^D$ to reconstruct $X$ from $Z$ by minimizing the negative log-likelihood (NLL) $-\mathbb{E}_{p(x,z)}[\log p_\theta(x|z)]$ of the conditional probability distribution that it parameterizes. Learning of the generator function $G_\phi: \mathbb{R}^K \mapsto \mathbb{R}^K$ which is referred to as \textit{learning the prior} in previous literature \citep{vqvae2,vqvae2o}, is essentially fitting a generative model to the marginal distribution of $Z$, i.e. $p(z)$. While we could use a similar approach as above and fit $G_\phi$ that parameterizes the model distribution $p_\phi(z)$ by minimizing $-\mathbb{E}_{p(z)}[\log p_\phi(z)]$, we chose to instead use a BigGAN model to directly minimize the (variational lower-bound to the) $f$-divergence $\mathbb{D}_f[p(z)\Vert p_\phi(z)]$ \citep{fgan} between the true marginal distribution $p(z)$ and the model distribution as it provides a better fit for natural-image like distributions compared to other approaches. All together, \wt simply specifies a generative model over $X$ and $Z$ jointly and is, therefore, trained by minimising the NLL,
\begin{equation}
    -\mathbb{E}_{p(x,z)}[\log p_{\theta, \phi}(x,z)] = -\mathbb{E}_{p(x,z)}[\log p_{\theta}(x|z)] -\mathbb{E}_{p(z)}[\log p_{\phi}(z)].
\end{equation}

\section{Experiments}

In this section, we quantitatively and qualitatively evaluate the training efficiency of the \wt approach against the end-to-end training approach of the BigGAN model by benchmarking the compute budget required vs. the quality of the generated samples. While quantifying image quality remains a challenging problem \citep{borji2019pros}, we report the Frechet Inception Distance (FID) \citep{fid} and Inception Score (IS) \citep{is} as proxy measures to allow for direct comparison with BigGAN. All results are reported on the ImageNet dataset \citep{imagenet} at two different resolutions, 256$\times$256 and 512$\times$512.

\paragraph{Compute Budget.} For our main experiments on training efficiency, depending on the sampler used, we define two different compute budgets. For the learned samplers, we train the \wtw and \wtp models for a total of 168 hours (7 full days). Our training is performed on a single machine with four Telsa V100 GPUs with 16GB of VRAM each. 
We found that training the baseline model (BigGAN) at 256x with only 4 GPUs in 7 days is not possible on the default setting, so we allow a total of 8 GPUs for the baseline training.
Since the training of these BigGAN-based samplers are highly unstable \citep{fqgan}, we train 5 instances in total (3 BigGAN-deep and 2 BigGAN samplers) and report results using the instance with the best FID score. For the pre-trained sampelrs, the compute budget is computed in terms of the TPUs used in their original training and does not include the decoder training. This is because the compute budget required to train the decoder is negligible compared to that of the sampler.

\paragraph{Hyperparameters and Setup.} For the deterministic encoder using wavelet transformation, we instantiate it with biorthogonal 2.2 wavelet. 
For the \wt sampler, we use a batch size of 512 and learning rates of $10^{-4}$ and $4 \times 10^{4}$ for the generator and discriminator, respectively. 
For the pre-training of ESRGAN-W and ESRGAN-P with L1 loss, we use a batch size of 32, and the learning rate is initialized $1 \times 10^{-4}$ and decayed by a factor of 2 every $2 \times 10^{5}$ mini-batch updates. During adversarial training of ESRGAN-W and ESRGAN-P, the learning rate is set to $1 \times 10^{-4}$ and halved at $[50k, 100k]$ iterations.
In the case of the UNet-decoders, for the first-level decoder, we use a batch size of $128$ and a learning rate of $10^{-4}$. For the second-level decoder, we use the same learning rate as the first-level decoder, but a smaller batch size of $64$. 
We allocate two GPUs to the \wt sampler, and the other two for the decoders. For the UNet-decoder, this implies one GPU per level. Each component is trained in parallel and independently from each other. To train the BigGAN baseline model at the native resolution of $256 \times 256$, we use the same learning rates as for the \wt sampler and a recommended batch size of $2048$ \citep{biggan}. The baseline model trains with eight GPUs, instead of four like with \wt, to meet its large memory requirement, given the batch size and resolution of the image. We provide training and evaluation code for \wt sampler and the other models, respectively here: \url{https://anonymous.4open.science/r/ca7a3f2e-5c27-48bd-a3bc-2dceadc138c1/}.


\subsection{Results}
\begin{table}[]
\centering
\begin{tabular}{@{}c|c|c|c|c@{}}
\toprule
 \textbf{Sampler} &  \textbf{Decoder}& \textbf{Resolution} & \textbf{min FID/IS} & \textbf{Compute Budget}\\ \midrule
Learned-P-64 & ESRGAN-P & 256 & 32.66 / 89.81 & 8 days on 4 $\times$ V100 GPUs\\ \midrule
\textbf{Learned-W-64} & \textbf{ESRGAN-W} & \textbf{256} & \textbf{21.82} / \textbf{119.8} & \textbf{7 days on 4 $\times$ V100 GPUs} \\ \midrule
Learned Baseline & None & 256 & $>$ 200 / $<$ 10  & 7 days on 8 $\times$ V100 GPUs \\ \bottomrule
\end{tabular}
\caption{\label{tab:learned} The learned samplers above are all realized with BigGAN/BigGAN-deep architecture. On a small compute budget, it is not only possible to train the Learned-W-64 sampler well, but also the \wtw model (with Learned-W-64 sampler) beats the state-of-the-art VQVAE-2 model (before truncation/rejection sampling), which has an extremely high cost of training similar to that of the BigGAN model. We defer a more detailed discussion about truncation and rejection sampling in Appendix \ref{sec:learned_samplers_detailed} and showcase a more detailed set of results in Table \ref{tab:learned_detailed} in the Appendix \ref{sec:learned_samplers_detailed}.}
\end{table}

\subsubsection{Learned Samplers}
To establish the training efficiency and competitive image quality of our \wt approach, we compare the FID and IS that the \wtw, \wtp and baseline BigGAN models reach in the compute budget of 7 days with four (8 for the baseline) Tesla V100 GPUs in Table \ref{tab:learned}. Both \wtp and \wtw models reach very competitive FIDs of 32.66 and 21.82, respectively. To put these results in perspective, even SAGAN \citep{sagan}, which requires twice the compute and operates on half the resolution (128 $\times$ 128), reaches an FID of $\sim$19. Furthermore, in the same amount of time but twice the compute, the baseline BigGAN model fails to generate any meaningful samples across all five runs.
This illustrates that, compared to end-to-end models, the \wt models are significantly more compute efficient and can reach competitive image quality.

In the aforementioned models, we use the public PyTorch implementation of BigGAN (kindly provided by the authors) to provide fair comparisons between the \wt models and the baselines. It is important to note, however, that the Pytorch implementation is not optimal as it does not use Sync-BatchNorm (SBN) which is essential for large mini-batch training. This implementation has been reported to not reach the FID and IS reported in the original paper \citep{brock_2018}. 

When the sampler is trained in the low-dimensional space of $64 \times 64$, \wtw clearly outperforms \wtp. We posit that this sharp difference in FID stems from the difference in the loss of structural information using the two down-sampling methods: pixel-based interpolation and wavelet-space encoding \citep{6922406}. As illustrated in Figure \ref{fig:interpol_comp} in Appendix \ref{app:a}, WT encoding seems to preserve more structural information than pixel-based interpolation which misses key features of the image due to arbitrary sub-sampling. This finding suggests that the down-sampled distribution at $64 \times 64$ becomes sufficiently different from the original distribution at $256 \times 256$ such that when trained on the down-sampled distribution, \wtp sampler fails to sufficiently approximate the structure in the original data distribution. If this were true, the \wtp performance should improve as the level of down-sampling is decreased. With the following set of experiments with pre-trained samplers, we test this hypothesis and demonstrate that \wt framework can also be used with pre-trained samplers to make image generation at high resolution ($256 \times 256, 512 \times 512$) efficient and even beat the BigGAN baseline model in some cases.

\subsubsection{Pre-trained Samplers}

\begin{table}[t]
\centering
\begin{tabular}{@{}c|c|c|c|c@{}}
\toprule
 \textbf{Sampler} &  \textbf{Decoder}& \textbf{Resolution} & \textbf{min FID / IS} & \textbf{Compute}\\ \midrule
BigGAN-128 & None & 128 & 10.58 / 43.72 & 128 TPUs \\ \midrule
BigGAN-256 & None & 256 & 10.68 / 52.16 & 256 TPUs \\ \midrule
Pretrained-128-64 & ESRGAN-P & 256 & 12.28 / 46.06 & 128 TPUs \\ \midrule
Pretrained-128-64 & ESRGAN-W & 256 & 12.66 / 45.54 & 128 TPUs \\ \midrule
BigGAN-512 & None & 512 & 11.32 / 49.37 & 512 TPUs \\ \midrule
\textbf{Pretrained-256-128} & \textbf{ESRGAN-P} & \textbf{512} & \textbf{10.30 / 213.4} & \textbf{256 TPUs} \\ \midrule
\textbf{Pretrained-256-128} & \textbf{ESRGAN-W} & \textbf{512} & \textbf{10.59 / 52.14} & \textbf{256 TPUs} \\ \midrule
Pretrained-128 & ESRGAN-P & 512 & 13.55 / 47.70 & 128 TPUs \\ \bottomrule
\end{tabular}
\caption{\label{tab:pretrained}All pretrained samplers above are realized with BigGAN. In higher dimensions, compared to the baseline BigGAN models at the respective resolutions, \wt models reduce the training compute budget by up to four times while suffering only a minor increase in FID. All pre-trained samplers above are trained for approximately two days on the compute described. Note that the decoder is only trained once, but generalizes across all the resolutions. This amortization further reduces the training cost drastically. Pretrained-128-64, for example, indicates that the model generates at $128 \times 128$ resolution and we down-sample it to $64 \times 64$ resolution with our encoders for \wt models. As empirically tested and confirmed by \cite{vqvae2}, IS is highly sensitive to slight blurs and perturbations. Therefore, we include an expanded set of quantitative results with various truncation and rejection sampling levels in Table \ref{tab:pretrained_detailed} in the Appendix \ref{sec:pretrained_sampler_detailed}.}
\end{table}
 We consider two pairs of pre-trained BigGAN samplers combined with our \wt models to ultimately generate at two different resolutions of 256$\times$256 and 512$\times$512. To clarify, our pre-trained samplers are simply a serial combination of pre-trained BigGAN models \footnote{We use the pre-trained models from huggingface's PyTorch re-implementation of BigGAN model.} and our encoders. For example, Pretrained-128-64 is created by generating samples from a pre-trained sampler at $128 \times 128$ and then applying a down-sampling operation (pixel or wavelet) to obtain $64 \times 64$ samples. Then, our decoders up-scale the low-resolution image four times to $256 \times 256$. We report the FID/IS scores for all the samplers and compare to the baseline BigGAN models pre-trained at the respective resolutions in Table \ref{tab:pretrained}. First, notice that \wtp performance increases drastically, supporting our hypothesis. Next, both \wtw and \wtp models require significantly less compute (up to 4 times less TPU-v3 cores) to train. Most importantly, note how approach outperforms the original BigGAN model in terms of FID at 512$\times$512 resolution, despite using exactly half the original compute budget. We hypothesize that this is partly due to the way in which ImageNet 512$\times$512 dataset is generated, but defer this discussion to Appendix \ref{sec:imagenet512}. 
 
 Furthermore, note that we did not have to re-train the decoders when applying to different resolutions and samplers. All results in this experiment are done with the decoders from the previous experiment that were only trained to up-scale images from $64 \times 64$ to $256 \times 256$. They evidently generalize well across different resolutions. This amortization of training across resolutions further reduces the training cost when more than one generator is trained. 

\begin{table}[]
\centering 
\begin{tabular}{@{}cccc@{}}
\toprule
 & \textbf{UNet-P} & \textbf{UNet-W} & \textbf{VQVAE-2} \\ \midrule
\textbf{Train MSE} & 0.0061 & \textbf{0.0045} & 0.0047 \\ \midrule
\textbf{Valid MSE} & 0.0074 & \textbf{0.0049} & 0.0050 \\ \bottomrule

\end{tabular}
\caption{\label{tab:mse}MSE on training and validation set for UNet-P, UNet-W,and VQVAE-2 models. Small difference between the training and validation error suggests that the models generalize well.}
\end{table}

\paragraph{UNet-Decoders}
As shown in Table \ref{tab:learned_detailed} of the Appendix \ref{sec:learned_samplers_detailed}, our ESRGAN-based decoders clearly outperform our UNet-decoders on FID and IS. It is important to note, however, that this difference is primarily because the ESRGAN decoders employ adversarial training which has been shown to drastically improve image quality as measured by FID. As shown in Table \ref{tab:mse}, not considering adversarial decoder, our UNet-W decoder leads to the best reconstruction error compared to the pixel-space decoder, including the state-of-the-art VQVAE-2 model.  

\paragraph{Evaluation on LSUN Church with StyleGAN-2-W}
The \wt approach can be implemented with models other than BigGAN. To demonstrate this, we replace the BigGAN sampler architecture in our model with the StyleGAN-2 architecture and re-run the experiments on the LSUN-Church dataset. The results in Table \ref{tab:style} of the Appendix \ref{sec:stylegan} demonstrate that our \wt models can generalize to different samplers and datasets. 

\paragraph{Qualitative Results.} Due to space limitations, we have deferred all qualitative results to the Appendix. Please see the Appendix \ref{sec:wt_learned_samples}, \ref{sec:wt_pretrained_samples}, \ref{sec:wtw_full_res}, and \ref{sec:wtp_full_res} for a detailed qualitative comparison of all the models.

\section{Conclusion}  
In this work, we present a new genre of compute-efficient generative models, \wtf, that achieve comparable image quality to the current SoTA DGM (BigGAN) with a dramatically lower compute budget. Surprisingly, \wtf is even able to outperform BigGAN in image quality at $512 \times 512$ resolution. Overall, we hope that our work inspires others to develop low-compute generative models that can be utilized and iterated on by the wider research community.


\clearpage

\bibliography{iclr2021_conference}

\begin{thebibliography}{46}
\providecommand{\natexlab}[1]{#1}
\providecommand{\url}[1]{\texttt{#1}}
\expandafter\ifx\csname urlstyle\endcsname\relax
  \providecommand{\doi}[1]{doi: #1}\else
  \providecommand{\doi}{doi: \begingroup \urlstyle{rm}\Url}\fi

\bibitem[ajbrock(2018)]{brock_2018}
ajbrock.
\newblock Training results(is and fid) are not good as yours with same training
  process · issue 23 · ajbrock/biggan-pytorch, 2018.
\newblock URL \url{https://github.com/ajbrock/BigGAN-PyTorch/issues/23}.

\bibitem[Antonini et~al.(1992)Antonini, Barlaud, Mathieu, and Daubechies]{wt1}
Marc Antonini, Michel Barlaud, Pierre Mathieu, and Ingrid Daubechies.
\newblock Image coding using wavelet transform.
\newblock \emph{IEEE Transactions on image processing}, 1\penalty0
  (2):\penalty0 205--220, 1992.

\bibitem[Arjovsky et~al.(2017)Arjovsky, Chintala, and Bottou]{wgan}
Martin Arjovsky, Soumith Chintala, and L{\'e}on Bottou.
\newblock Wasserstein gan.
\newblock \emph{arXiv preprint arXiv:1701.07875}, 2017.

\bibitem[Borji(2019)]{borji2019pros}
Ali Borji.
\newblock Pros and cons of gan evaluation measures.
\newblock \emph{Computer Vision and Image Understanding}, 179:\penalty0 41--65,
  2019.

\bibitem[Brock et~al.(2018)Brock, Donahue, and Simonyan]{biggan}
Andrew Brock, Jeff Donahue, and Karen Simonyan.
\newblock Large scale gan training for high fidelity natural image synthesis.
\newblock In \emph{International Conference on Learning Representations}, 2018.

\bibitem[Chen et~al.(2017)Chen, Mishra, Rohaninejad, and Abbeel]{pixelsnail}
Xi~Chen, Nikhil Mishra, Mostafa Rohaninejad, and Pieter Abbeel.
\newblock Pixelsnail: An improved autoregressive generative model.
\newblock \emph{arXiv preprint arXiv:1712.09763}, 2017.

\bibitem[Daubechies(1992)]{wt}
Ingrid Daubechies.
\newblock \emph{Ten lectures on wavelets}, volume~61.
\newblock Siam, 1992.

\bibitem[De~Fauw et~al.(2019)De~Fauw, Dieleman, and Simonyan]{vqvae2o}
Jeffrey De~Fauw, Sander Dieleman, and Karen Simonyan.
\newblock Hierarchical autoregressive image models with auxiliary decoders.
\newblock \emph{arXiv preprint arXiv:1903.04933}, 2019.

\bibitem[{Deng} et~al.(2009){Deng}, {Dong}, {Socher}, {Li}, {Kai Li}, and {Li
  Fei-Fei}]{imagenet}
J.~{Deng}, W.~{Dong}, R.~{Socher}, L.~{Li}, {Kai Li}, and {Li Fei-Fei}.
\newblock Imagenet: A large-scale hierarchical image database.
\newblock In \emph{2009 IEEE Conference on Computer Vision and Pattern
  Recognition}, pp.\  248--255, 2009.

\bibitem[Denton et~al.(2015)Denton, Chintala, Fergus, et~al.]{denton2015deep}
Emily~L Denton, Soumith Chintala, Rob Fergus, et~al.
\newblock Deep generative image models using a laplacian pyramid of adversarial
  networks.
\newblock In \emph{Advances in neural information processing systems}, pp.\
  1486--1494, 2015.

\bibitem[Dong et~al.(2014)Dong, Loy, He, and Tang]{srcnn}
Chao Dong, Chen~Change Loy, Kaiming He, and Xiaoou Tang.
\newblock Learning a deep convolutional network for image super-resolution.
\newblock In David Fleet, Tomas Pajdla, Bernt Schiele, and Tinne Tuytelaars
  (eds.), \emph{Computer Vision -- ECCV 2014}, pp.\  184--199, Cham, 2014.
  Springer International Publishing.
\newblock ISBN 978-3-319-10593-2.

\bibitem[Goodfellow et~al.(2014)Goodfellow, Pouget{-}Abadie, Mirza, Xu,
  Warde{-}Farley, Ozair, Courville, and Bengio]{gan}
Ian~J. Goodfellow, Jean Pouget{-}Abadie, Mehdi Mirza, Bing Xu, David
  Warde{-}Farley, Sherjil Ozair, Aaron~C. Courville, and Yoshua Bengio.
\newblock Generative adversarial nets.
\newblock In \emph{Neural Information Processing Systems}, 2014.

\bibitem[Gulrajani et~al.(2017)Gulrajani, Ahmed, Arjovsky, Dumoulin, and
  Courville]{wgangp}
Ishaan Gulrajani, Faruk Ahmed, Martin Arjovsky, Vincent Dumoulin, and Aaron~C
  Courville.
\newblock Improved training of wasserstein gans.
\newblock In \emph{Advances in neural information processing systems}, pp.\
  5767--5777, 2017.

\bibitem[Haar(1909)]{wt0}
Alfred Haar.
\newblock \emph{on the theory of orthogonal functional systems}.
\newblock Georg August University, Gottingen., 1909.

\bibitem[He et~al.(2016)He, Zhang, Ren, and Sun]{resnet}
Kaiming He, X.~Zhang, Shaoqing Ren, and Jian Sun.
\newblock Deep residual learning for image recognition.
\newblock \emph{2016 IEEE Conference on Computer Vision and Pattern Recognition
  (CVPR)}, pp.\  770--778, 2016.

\bibitem[Heusel et~al.(2017)Heusel, Ramsauer, Unterthiner, Nessler, and
  Hochreiter]{fid}
Martin Heusel, Hubert Ramsauer, Thomas Unterthiner, Bernhard Nessler, and Sepp
  Hochreiter.
\newblock {GANS} trained by a two time-scale update rule converge to a local
  nash equilibrium.
\newblock In \emph{Neural Information Processing Systems}, 2017.

\bibitem[Hu et~al.(2019)Hu, Naiel, Wong, Lamm, and Fieguth]{hu2019runet}
Xiaodan Hu, Mohamed~A Naiel, Alexander Wong, Mark Lamm, and Paul Fieguth.
\newblock Runet: A robust unet architecture for image super-resolution.
\newblock In \emph{Proceedings of the IEEE Conference on Computer Vision and
  Pattern Recognition Workshops}, pp.\  0--0, 2019.

\bibitem[{Jo} et~al.(2018){Jo}, {Oh}, {Kang}, and {Kim}]{duf}
Y.~{Jo}, S.~W. {Oh}, J.~{Kang}, and S.~J. {Kim}.
\newblock Deep video super-resolution network using dynamic upsampling filters
  without explicit motion compensation.
\newblock In \emph{2018 IEEE/CVF Conference on Computer Vision and Pattern
  Recognition}, pp.\  3224--3232, 2018.

\bibitem[Jouppi et~al.(2017)Jouppi, Young, Patil, Patterson, Agrawal, Bajwa,
  Bates, Bhatia, Boden, Borchers, et~al.]{tpu}
Norman~P Jouppi, Cliff Young, Nishant Patil, David Patterson, Gaurav Agrawal,
  Raminder Bajwa, Sarah Bates, Suresh Bhatia, Nan Boden, Al~Borchers, et~al.
\newblock In-datacenter performance analysis of a tensor processing unit.
\newblock In \emph{Proceedings of the 44th Annual International Symposium on
  Computer Architecture}, pp.\  1--12, 2017.

\bibitem[Karras et~al.(2017)Karras, Aila, Laine, and Lehtinen]{progressive}
Tero Karras, Timo Aila, Samuli Laine, and Jaakko Lehtinen.
\newblock Progressive growing of gans for improved quality, stability, and
  variation.
\newblock \emph{arXiv preprint arXiv:1710.10196}, 2017.

\bibitem[Karras et~al.(2020)Karras, Laine, Aittala, Hellsten, Lehtinen, and
  Aila]{stylegan}
Tero Karras, Samuli Laine, Miika Aittala, Janne Hellsten, Jaakko Lehtinen, and
  Timo Aila.
\newblock Analyzing and improving the image quality of stylegan.
\newblock In \emph{Proceedings of the IEEE/CVF Conference on Computer Vision
  and Pattern Recognition}, pp.\  8110--8119, 2020.

\bibitem[Kingma \& Welling(2013)Kingma and Welling]{vae}
Diederik~P Kingma and Max Welling.
\newblock Auto-encoding variational bayes.
\newblock \emph{arXiv preprint arXiv:1312.6114}, 2013.

\bibitem[Ledig et~al.(2017)Ledig, Theis, Husz{\'a}r, Caballero, Cunningham,
  Acosta, Aitken, Tejani, Totz, Wang, et~al.]{srgan}
Christian Ledig, Lucas Theis, Ferenc Husz{\'a}r, Jose Caballero, Andrew
  Cunningham, Alejandro Acosta, Andrew Aitken, Alykhan Tejani, Johannes Totz,
  Zehan Wang, et~al.
\newblock Photo-realistic single image super-resolution using a generative
  adversarial network.
\newblock In \emph{Proceedings of the IEEE conference on computer vision and
  pattern recognition}, pp.\  4681--4690, 2017.

\bibitem[Li et~al.(2017)Li, Chang, Cheng, Yang, and P{\'o}czos]{mmdgan}
Chun-Liang Li, Wei-Cheng Chang, Yu~Cheng, Yiming Yang, and Barnab{\'a}s
  P{\'o}czos.
\newblock Mmd gan: Towards deeper understanding of moment matching network.
\newblock In \emph{Advances in Neural Information Processing Systems}, pp.\
  2203--2213, 2017.

\bibitem[Lim et~al.(2017)Lim, Son, Kim, Nah, and Lee]{edsr}
Bee Lim, Sanghyun Son, Heewon Kim, Seungjun Nah, and Kyoung~Mu Lee.
\newblock Enhanced deep residual networks for single image super-resolution,
  2017.

\bibitem[Liu et~al.(2019{\natexlab{a}})Liu, Yao, and Ren]{acceleration}
Jinlin Liu, Yuan Yao, and Jianqiang Ren.
\newblock An acceleration framework for high resolution image synthesis,
  2019{\natexlab{a}}.

\bibitem[Liu et~al.(2019{\natexlab{b}})Liu, Yao, and Ren]{liu2019acceleration}
Jinlin Liu, Yuan Yao, and Jianqiang Ren.
\newblock An acceleration framework for high resolution image synthesis.
\newblock \emph{arXiv preprint arXiv:1909.03611}, 2019{\natexlab{b}}.

\bibitem[Menick \& Kalchbrenner(2019)Menick and Kalchbrenner]{spn}
Jacob Menick and Nal Kalchbrenner.
\newblock {GENERATING} {HIGH} {FIDELITY} {IMAGES} {WITH} {SUBSCALE} {PIXEL}
  {NETWORKS} {AND} {MULTIDIMENSIONAL} {UPSCALING}.
\newblock In \emph{International Conference on Learning Representations}, 2019.
\newblock URL \url{https://openreview.net/forum?id=HylzTiC5Km}.

\bibitem[Mescheder et~al.(2017)Mescheder, Nowozin, and Geiger]{lars2}
Lars Mescheder, Sebastian Nowozin, and Andreas Geiger.
\newblock The numerics of gans.
\newblock In \emph{Advances in Neural Information Processing Systems}, pp.\
  1825--1835, 2017.

\bibitem[Mescheder et~al.(2018)Mescheder, Geiger, and Nowozin]{lars1}
Lars Mescheder, Andreas Geiger, and Sebastian Nowozin.
\newblock Which training methods for gans do actually converge?
\newblock \emph{arXiv preprint arXiv:1801.04406}, 2018.

\bibitem[Miyato et~al.(2018)Miyato, Kataoka, Koyama, and Yoshida]{spectral}
Takeru Miyato, Toshiki Kataoka, Masanori Koyama, and Yuichi Yoshida.
\newblock Spectral normalization for generative adversarial networks.
\newblock In \emph{International Conference on Learning Representations}, 2018.
\newblock URL \url{https://openreview.net/forum?id=B1QRgziT-}.

\bibitem[Nowozin et~al.(2016)Nowozin, Cseke, and Tomioka]{fgan}
Sebastian Nowozin, Botond Cseke, and Ryota Tomioka.
\newblock f-gan: Training generative neural samplers using variational
  divergence minimization.
\newblock In \emph{Advances in neural information processing systems}, pp.\
  271--279, 2016.

\bibitem[Razavi et~al.(2019)Razavi, van~den Oord, and Vinyals]{vqvae2}
Ali Razavi, Aaron van~den Oord, and Oriol Vinyals.
\newblock Generating diverse high-fidelity images with vq-vae-2.
\newblock In \emph{Advances in Neural Information Processing Systems}, pp.\
  14837--14847, 2019.

\bibitem[Ronneberger et~al.(2015)Ronneberger, Fischer, and Brox]{unet}
Olaf Ronneberger, Philipp Fischer, and Thomas Brox.
\newblock U-net: Convolutional networks for biomedical image segmentation.
\newblock In \emph{International Conference on Medical image computing and
  computer-assisted intervention}, pp.\  234--241. Springer, 2015.

\bibitem[Roth et~al.(2017)Roth, Lucchi, Nowozin, and Hofmann]{kevin}
Kevin Roth, Aurelien Lucchi, Sebastian Nowozin, and Thomas Hofmann.
\newblock Stabilizing training of generative adversarial networks through
  regularization.
\newblock In \emph{Advances in neural information processing systems}, pp.\
  2018--2028, 2017.

\bibitem[Salimans et~al.(2016)Salimans, Goodfellow, Zaremba, Cheung, Radford,
  and Chen]{is}
Tim Salimans, Ian Goodfellow, Wojciech Zaremba, Vicki Cheung, Alec Radford, and
  Xi~Chen.
\newblock Improved techniques for training gans.
\newblock In \emph{Advances in neural information processing systems}, pp.\
  2234--2242, 2016.

\bibitem[Schwab(2018)]{env}
Katharine Schwab.
\newblock A google intern built the ai behind these shockingly good fake
  images, October 2018.
\newblock URL
  \url{https://www.fastcompany.com/90244767/see-the-shockingly-realistic-images-made-by-googles-new-ai}.
\newblock [Online; posted 10-02-18].

\bibitem[{Sekar} et~al.(2014){Sekar}, {Duraisamy}, and {Remimol}]{6922406}
K.~{Sekar}, V.~{Duraisamy}, and A.~M. {Remimol}.
\newblock An approach of image scaling using dwt and bicubic interpolation.
\newblock In \emph{2014 International Conference on Green Computing
  Communication and Electrical Engineering (ICGCCEE)}, pp.\  1--5, 2014.

\bibitem[Simonyan \& Zisserman(2015)Simonyan and Zisserman]{vgg}
Karen Simonyan and Andrew Zisserman.
\newblock Very deep convolutional networks for large-scale image recognition,
  2015.

\bibitem[Srivastava et~al.(2017)Srivastava, Valkov, Russell, Gutmann, and
  Sutton]{veegan}
Akash Srivastava, Lazar Valkov, Chris Russell, Michael~U Gutmann, and Charles
  Sutton.
\newblock Veegan: Reducing mode collapse in gans using implicit variational
  learning.
\newblock In \emph{Advances in Neural Information Processing Systems}, pp.\
  3308--3318, 2017.

\bibitem[Srivastava et~al.(2020)Srivastava, Xu, Gutmann, and Sutton]{gram}
Akash Srivastava, Kai Xu, Michael~U. Gutmann, and Charles Sutton.
\newblock Generative ratio matching networks.
\newblock In \emph{International Conference on Learning Representations}, 2020.
\newblock URL \url{https://openreview.net/forum?id=SJg7spEYDS}.

\bibitem[van~den Oord et~al.(2017)van~den Oord, Vinyals, et~al.]{vqvae}
Aaron van~den Oord, Oriol Vinyals, et~al.
\newblock Neural discrete representation learning.
\newblock In \emph{Advances in Neural Information Processing Systems}, pp.\
  6306--6315, 2017.

\bibitem[Wang et~al.(2018)Wang, Yu, Wu, Gu, Liu, Dong, Qiao, and
  Change~Loy]{esrgan}
Xintao Wang, Ke~Yu, Shixiang Wu, Jinjin Gu, Yihao Liu, Chao Dong, Yu~Qiao, and
  Chen Change~Loy.
\newblock Esrgan: Enhanced super-resolution generative adversarial networks.
\newblock In \emph{Proceedings of the European Conference on Computer Vision
  (ECCV)}, pp.\  0--0, 2018.

\bibitem[Zhang et~al.(2018{\natexlab{a}})Zhang, Goodfellow, Metaxas, and
  Odena]{sagan}
Han Zhang, Ian Goodfellow, Dimitris Metaxas, and Augustus Odena.
\newblock Self-attention generative adversarial networks.
\newblock \emph{arXiv preprint arXiv:1805.08318}, 2018{\natexlab{a}}.

\bibitem[Zhang et~al.(2018{\natexlab{b}})Zhang, Tian, Kong, Zhong, and Fu]{rdn}
Yulun Zhang, Yapeng Tian, Yu~Kong, Bineng Zhong, and Yun Fu.
\newblock Residual dense network for image super-resolution,
  2018{\natexlab{b}}.

\bibitem[Zhao et~al.(2020)Zhao, Li, Yu, Gao, and Chen]{fqgan}
Yang Zhao, Chunyuan Li, Ping Yu, Jianfeng Gao, and Changyou Chen.
\newblock Feature quantization improves gan training.
\newblock \emph{arXiv preprint arXiv:2004.02088}, 2020.

\end{thebibliography}
\bibliographystyle{iclr2021_conference}

\appendix
\section{Appendix}
\label{app:a}
\begin{figure}[htb!] 
\centering
\includegraphics[width=0.60\linewidth]{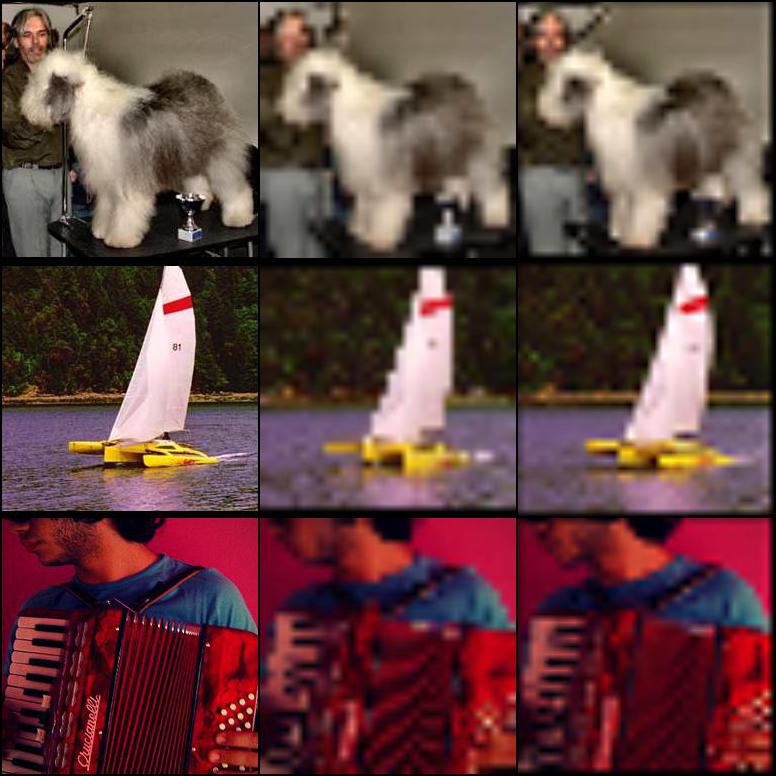}
\label{fig:downsampling}
\caption{\label{fig:interpol_comp} The above images are the original, down-sampled version using wavelet encoding, and down-sampled version using pixel-based interpolation (bilinear) in the column-wise order. Please note that the pixel-based interpolation loses key structural information in the images -- face of the person, structure of the boat, and accordion keyboard -- whereas wavelet encoding does not.}
\end{figure}

\subsection{\wt Samples with Learned Sampler}
\label{sec:wt_learned_samples}

\begin{figure}[t] 
\centering
\subfloat[\wtp with learned sampler]{
    \includegraphics[width=0.45\linewidth]{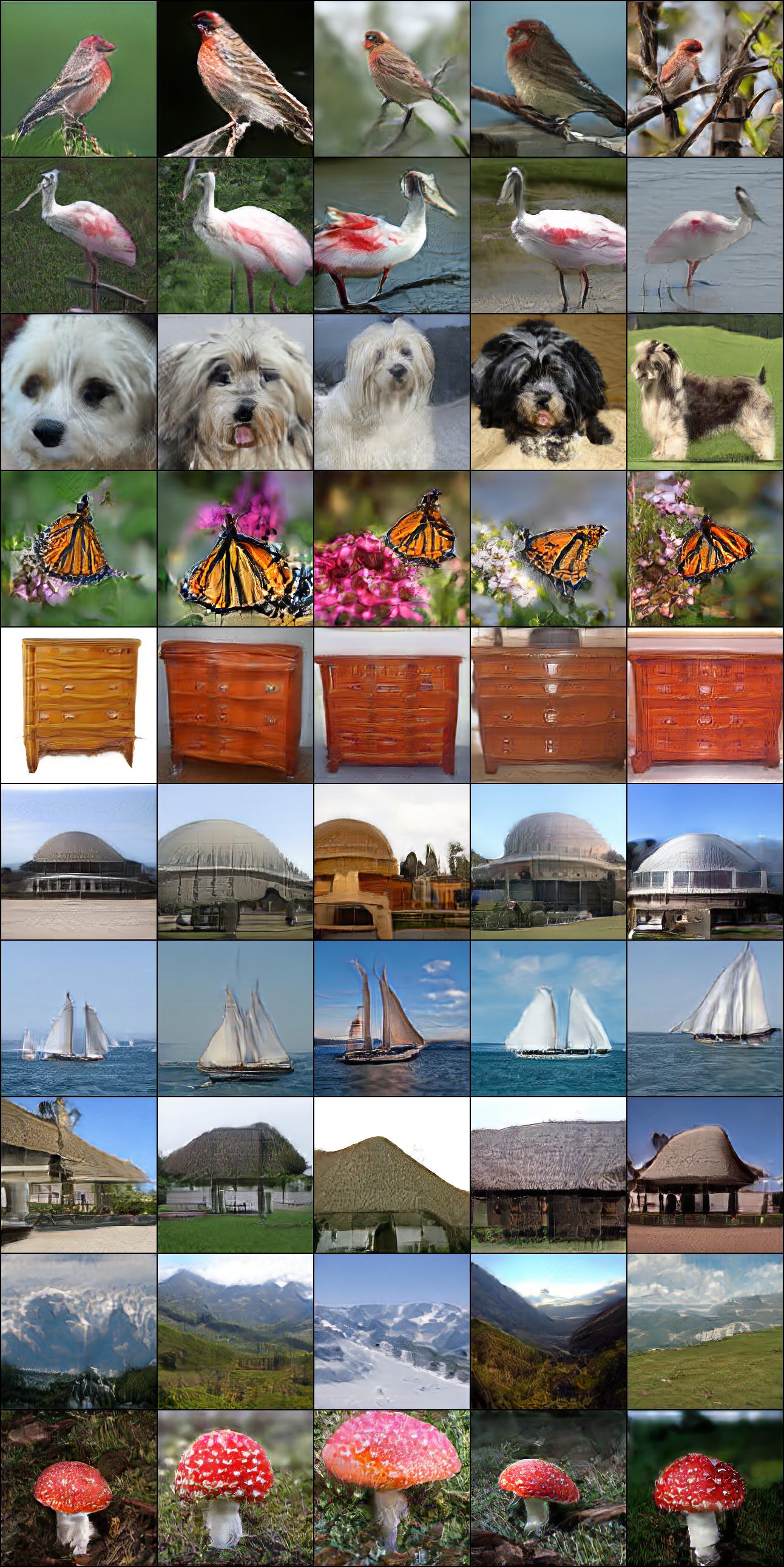}
    \label{fig:nsbgan_p_learned_samples}
}
\subfloat[\wtw with learned sampler]{
    \includegraphics[width=0.45\linewidth]{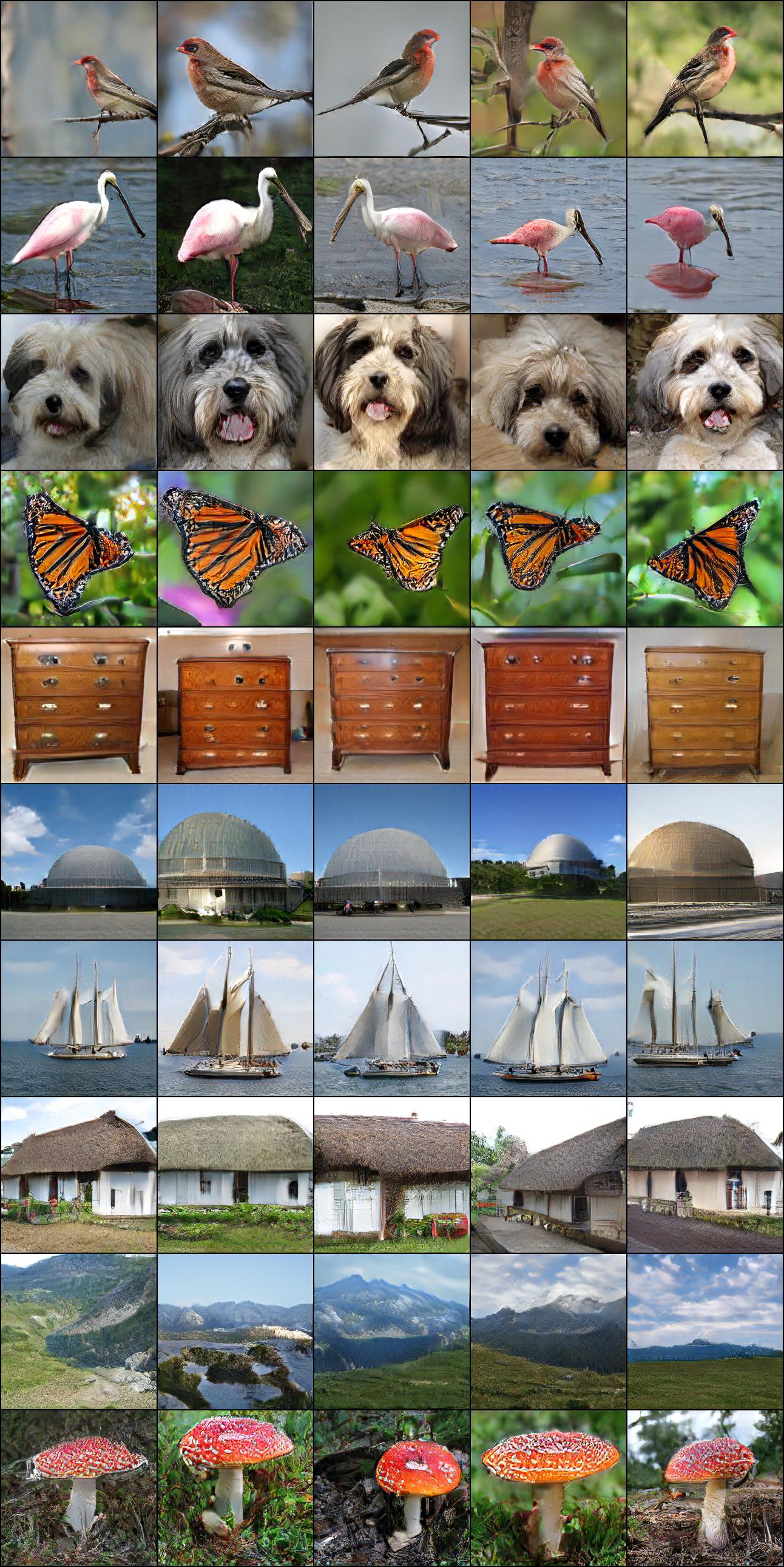}
    \label{fig:nsbgan_w_learned_samples}
}
\caption{\label{fig:nsbgan_learned_samples}Class-conditional random samples at $256 \times 256$ from \wt models with learned samplers. Classes from the top row: 12 house finch, 129 spoonbill, 200 Tibetan terrier, 323 monarch butterfly, 493 chiffonier, 727 planetarium, 780 schooner, 853 thatch, 970 alp, and 992 agaric.}
\end{figure}

Refer to Figure \ref{fig:nsbgan_learned_samples} for samples from \wt models.

\subsection{\wt Samples with Pre-trained Sampler}
\label{sec:wt_pretrained_samples}

\begin{figure}[t] 
\centering
\subfloat[\wtw with pre-trained sampler]{
    \includegraphics[width=0.45\linewidth]{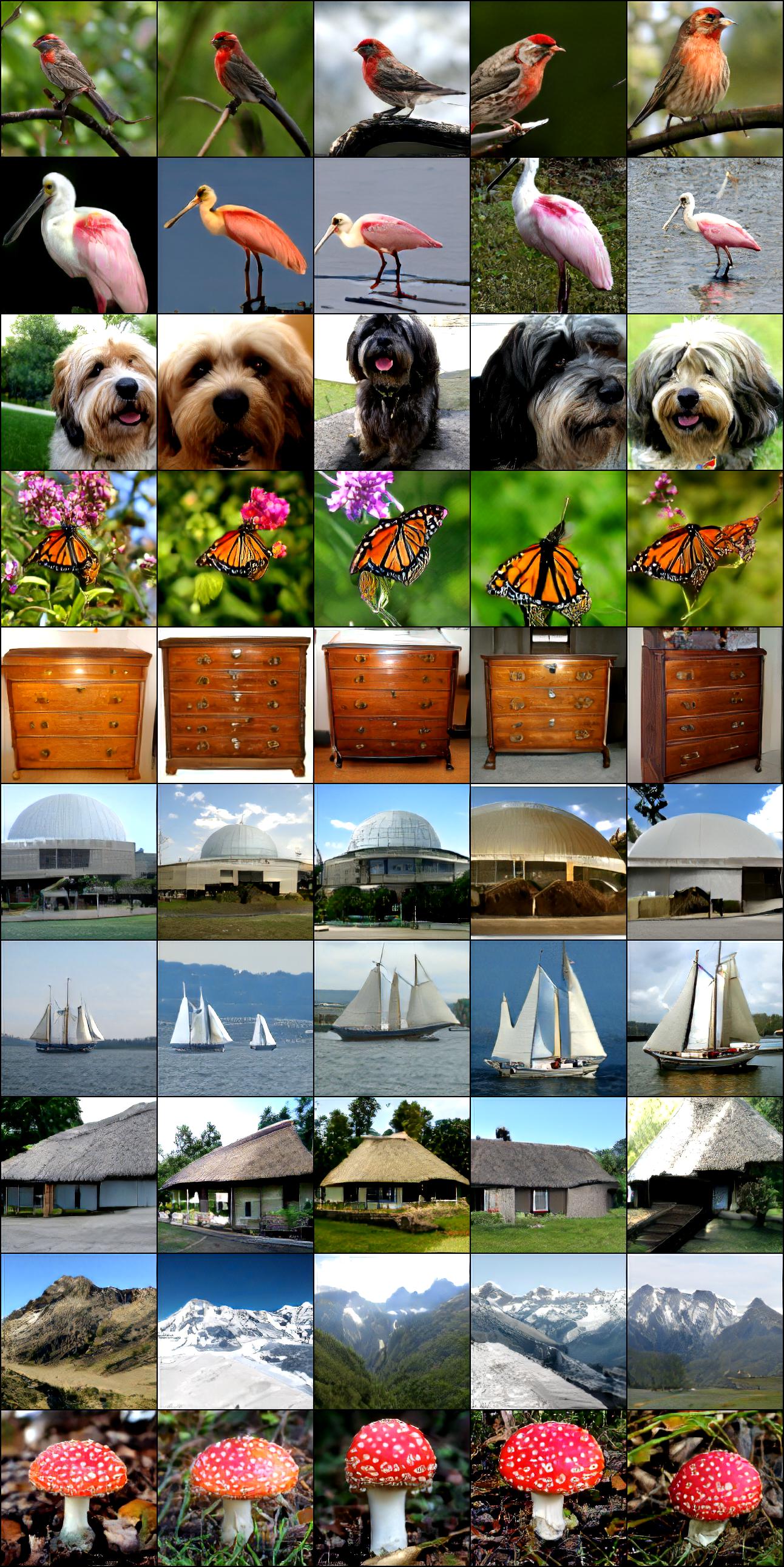}
    \label{fig:nsbgan_w_pretrained_samples_256}
}
\subfloat[\wtp with pre-trained sampler]{
    \includegraphics[width=0.45\linewidth]{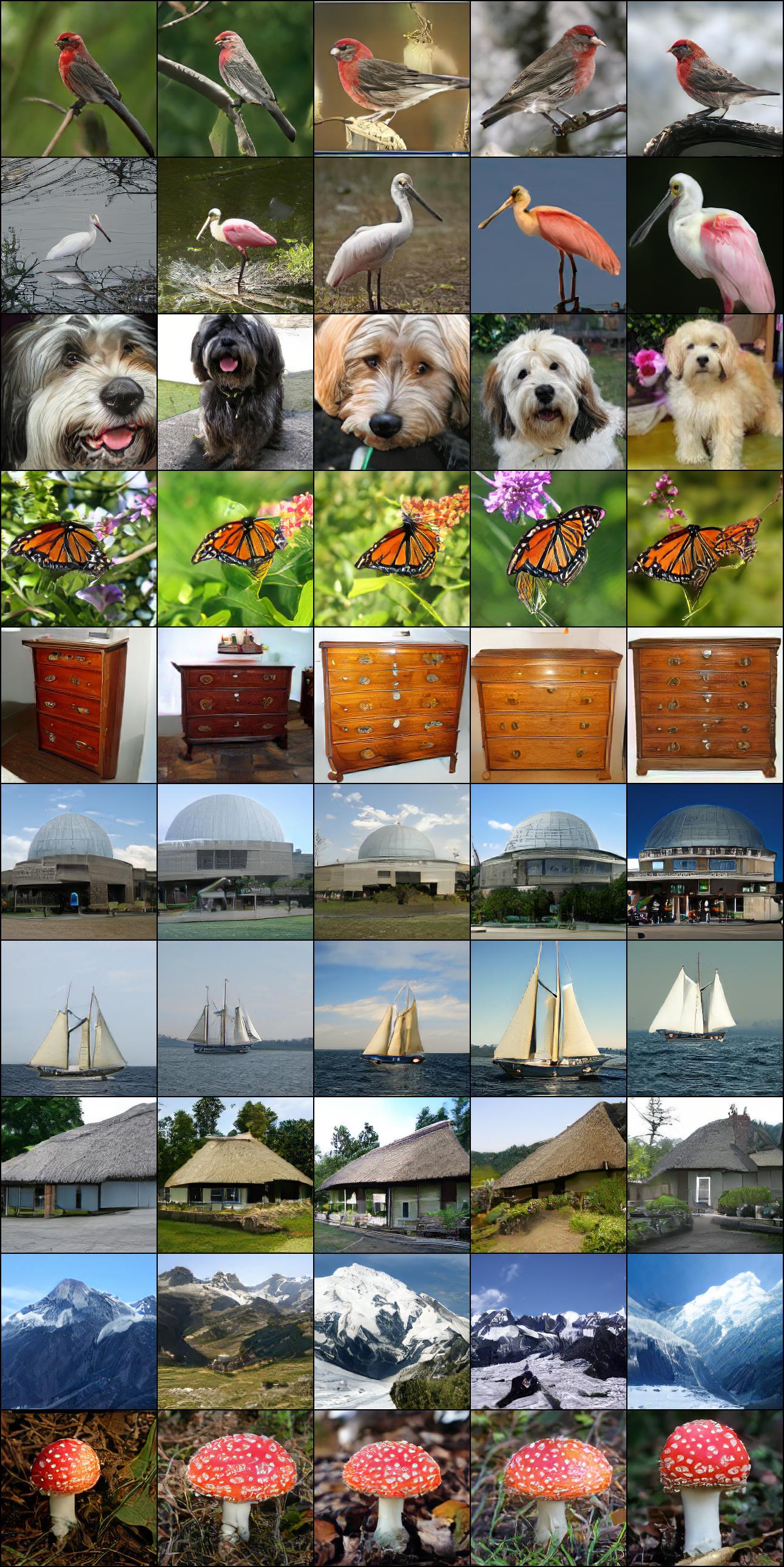}
    \label{fig:nsbgan_p_pretrained_samples_256}
}
\caption{Class-conditional random samples at $256 \times 256$ from \wt models with pre-trained samplers. Classes from the top row: 12 house finch, 129 spoonbill, 200 Tibetan terrier, 323 monarch butterfly, 493 chiffonier, 727 planetarium, 780 schooner, 853 thatch, 970 alp, and 992 agaric.}
\end{figure}

\begin{figure}[t] 
\centering
\subfloat[\wtw with pre-trained sampler]{
    \includegraphics[width=0.90\linewidth]{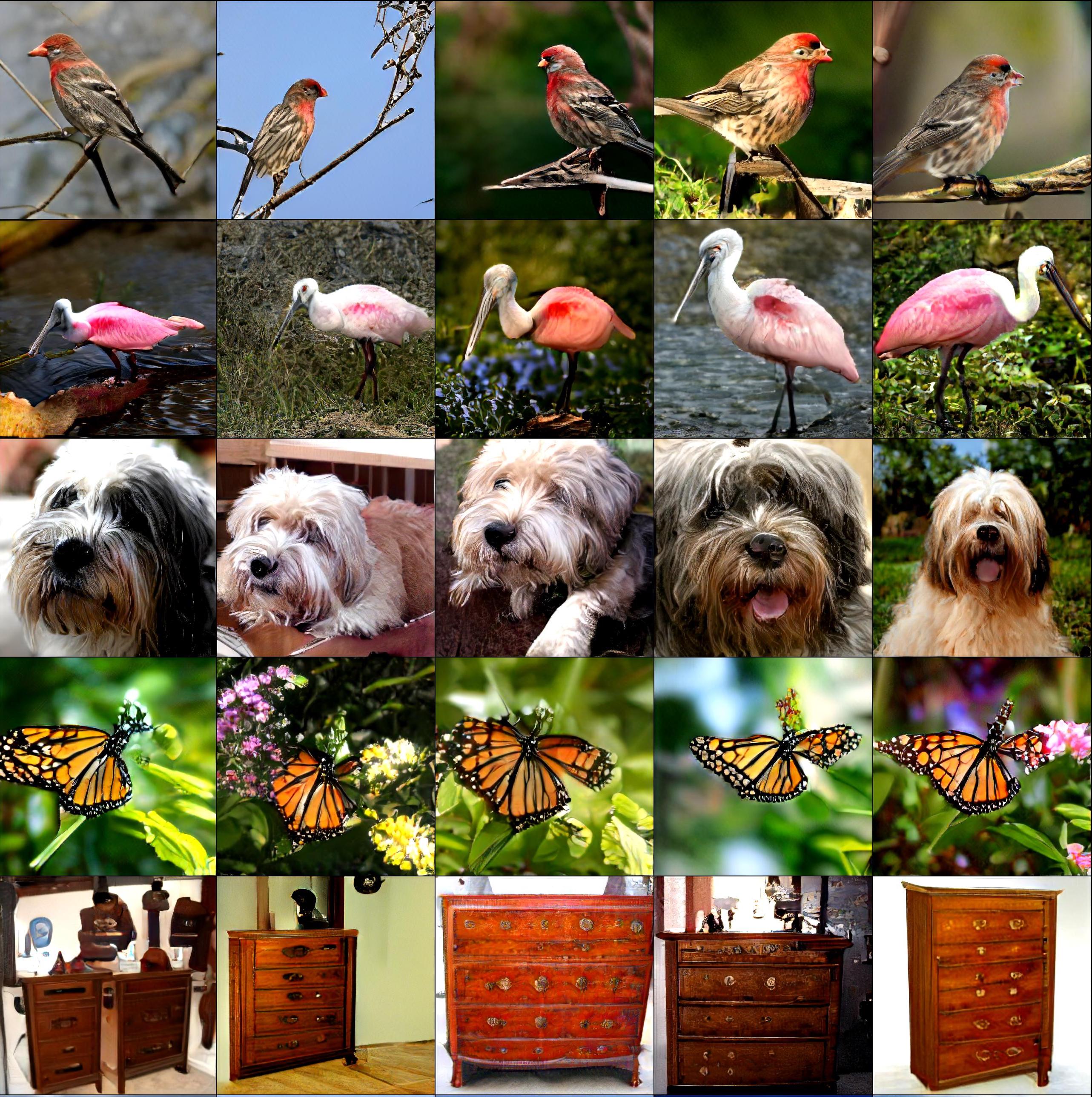}
    \label{fig:nsbgan_w_pretrained_samples_512_1}
}
\caption{Class-conditional random samples at $512 \times 512$ from \wtw model with pre-trained sampler. Classes from the top row: 12 house finch, 129 spoonbill, 200 Tibetan terrier, 323 monarch butterfly, and 493 chiffonier.}
\end{figure}

\begin{figure}[t] 
\centering
\subfloat[\wtw with pre-trained sampler]{
    \includegraphics[width=0.90\linewidth]{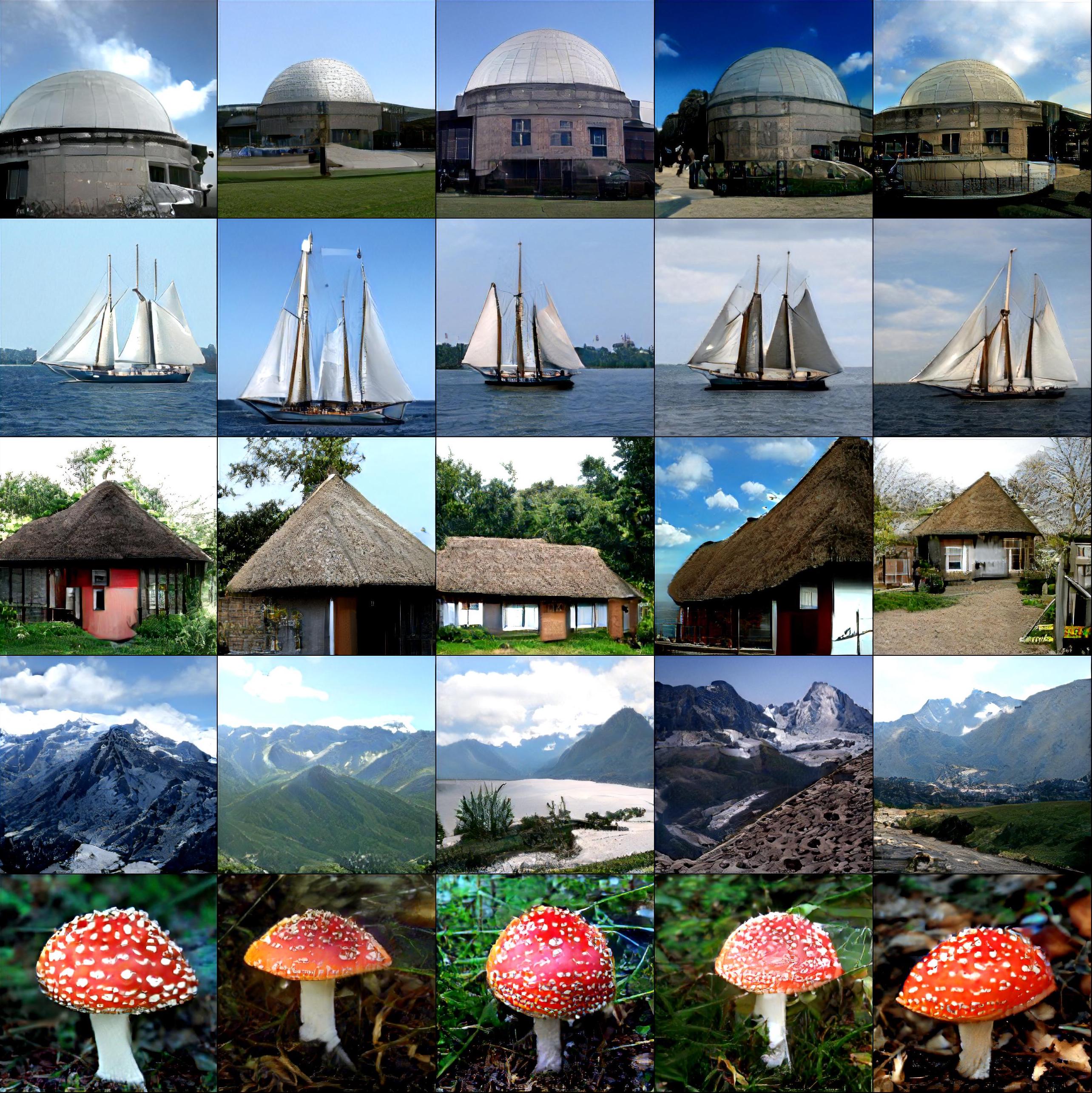}
    \label{fig:nsbgan_w_pretrained_samples_512_2}
}
\caption{Additional class-conditional random samples at $512 \times 512$ from \wtw model with pre-trained sampler. Classes from the top row: 727 planetarium, 780 schooner, 853 thatch, 970 alp, and 992 agaric.}
\end{figure}

\begin{figure}[t] 
\centering
\subfloat[\wtw with pre-trained sampler]{
    \includegraphics[width=0.90\linewidth]{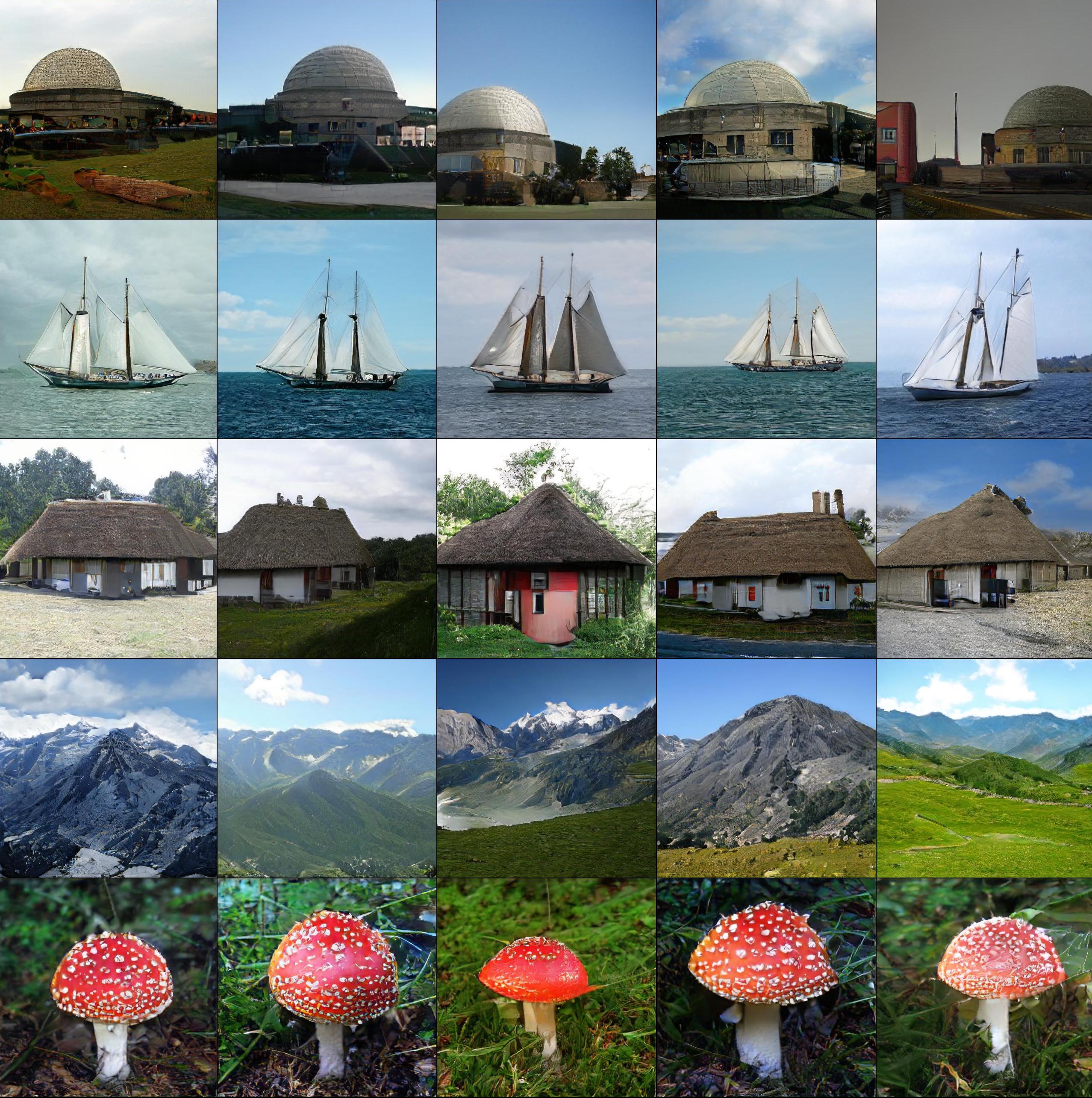}
    \label{fig:nsbgan_p_pretrained_samples_512_1}
}
\caption{Class-conditional random samples at $512 \times 512$ from \wtp model with pre-trained sampler. Classes from the top row: 12 house finch, 129 spoonbill, 200 Tibetan terrier, 323 monarch butterfly, and 493 chiffonier.}
\end{figure}

\begin{figure}[t] 
\centering
\subfloat[\wtw with pre-trained sampler]{
    \includegraphics[width=0.90\linewidth]{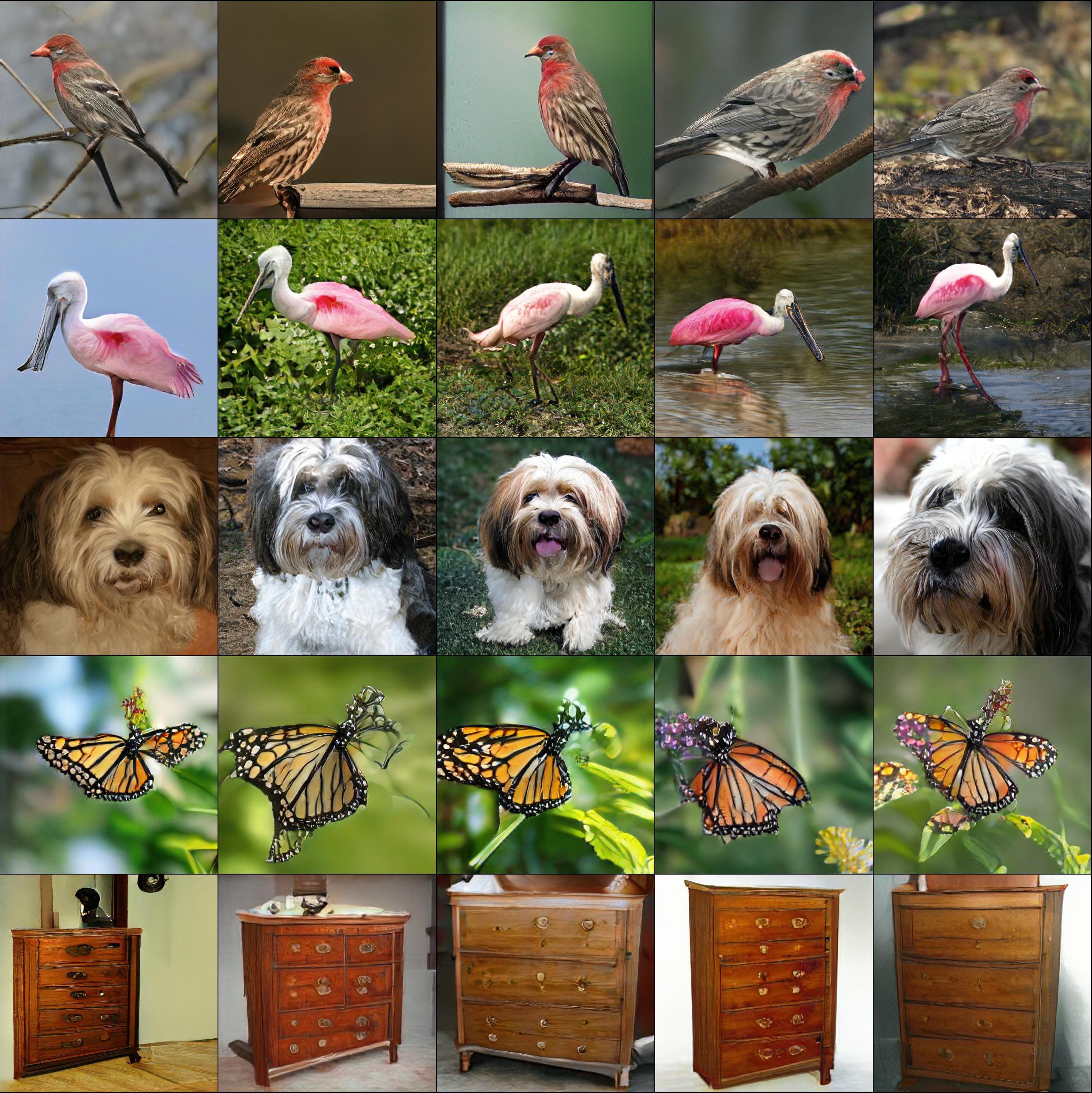}
    \label{fig:nsbgan_p_pretrained_samples_512_2}
}
\caption{Additional class-conditional random samples at $512 \times 512$ from \wtp model with pre-trained sampler. Classes from the top row: 12 house finch, 129 spoonbill, 200 Tibetan terrier, 323 monarch butterfly, and 493 chiffonier.}
\end{figure}

\begin{figure}[t] 
\centering
\includegraphics[width=0.60\linewidth]{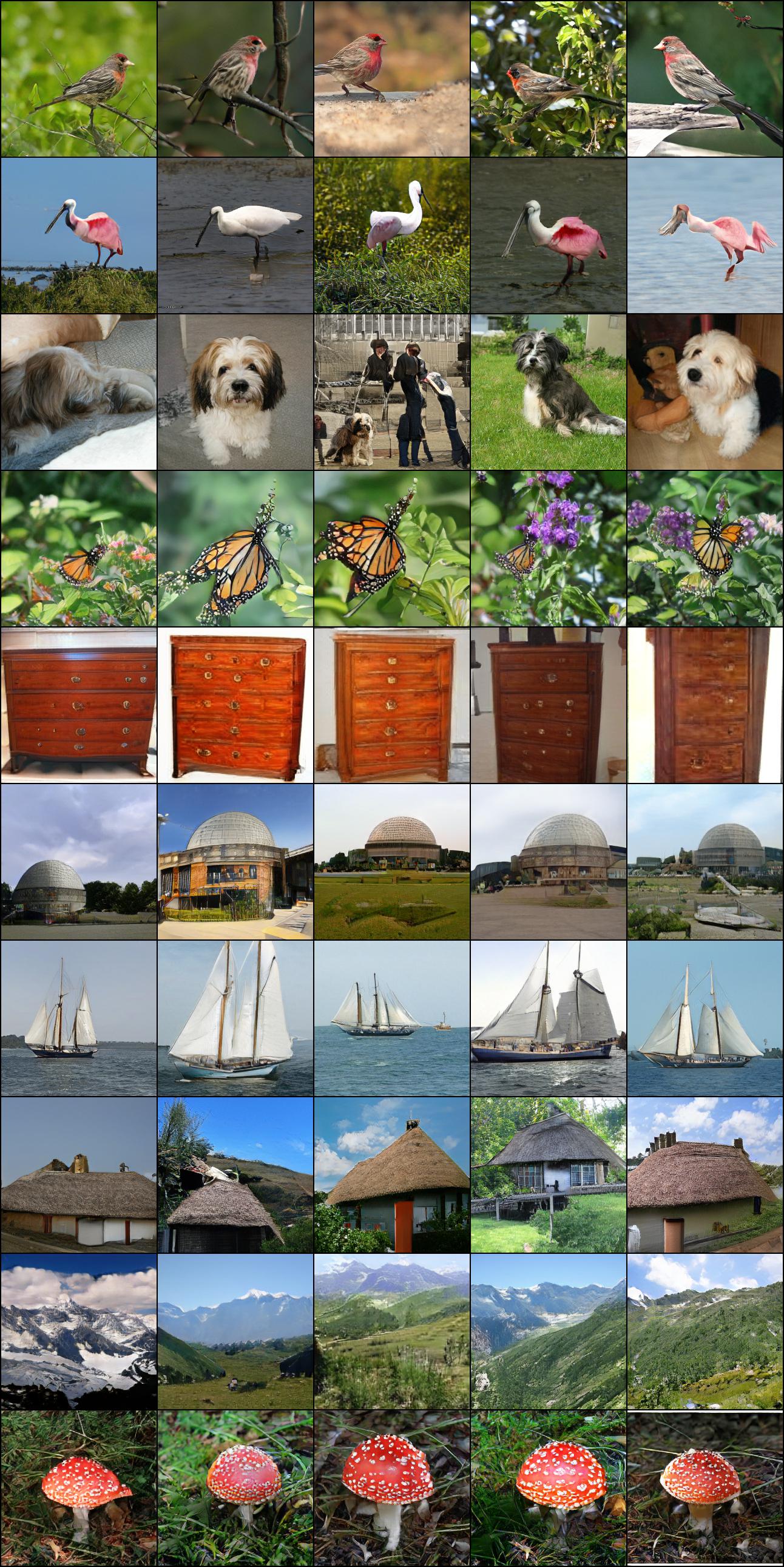}
\caption{\label{fig:biggan_256_samples}Class-conditional random samples at $256 \times 256$ from BigGAN. Classes from the top row: 12 house finch, 129 spoonbill, 200 Tibetan terrier, 323 monarch butterfly, 493 chiffonier, 727 planetarium, 780 schooner, 853 thatch, 970 alp, and 992 agaric.}
\end{figure}

\begin{figure}[t] 
\centering
\includegraphics[width=0.90\linewidth]{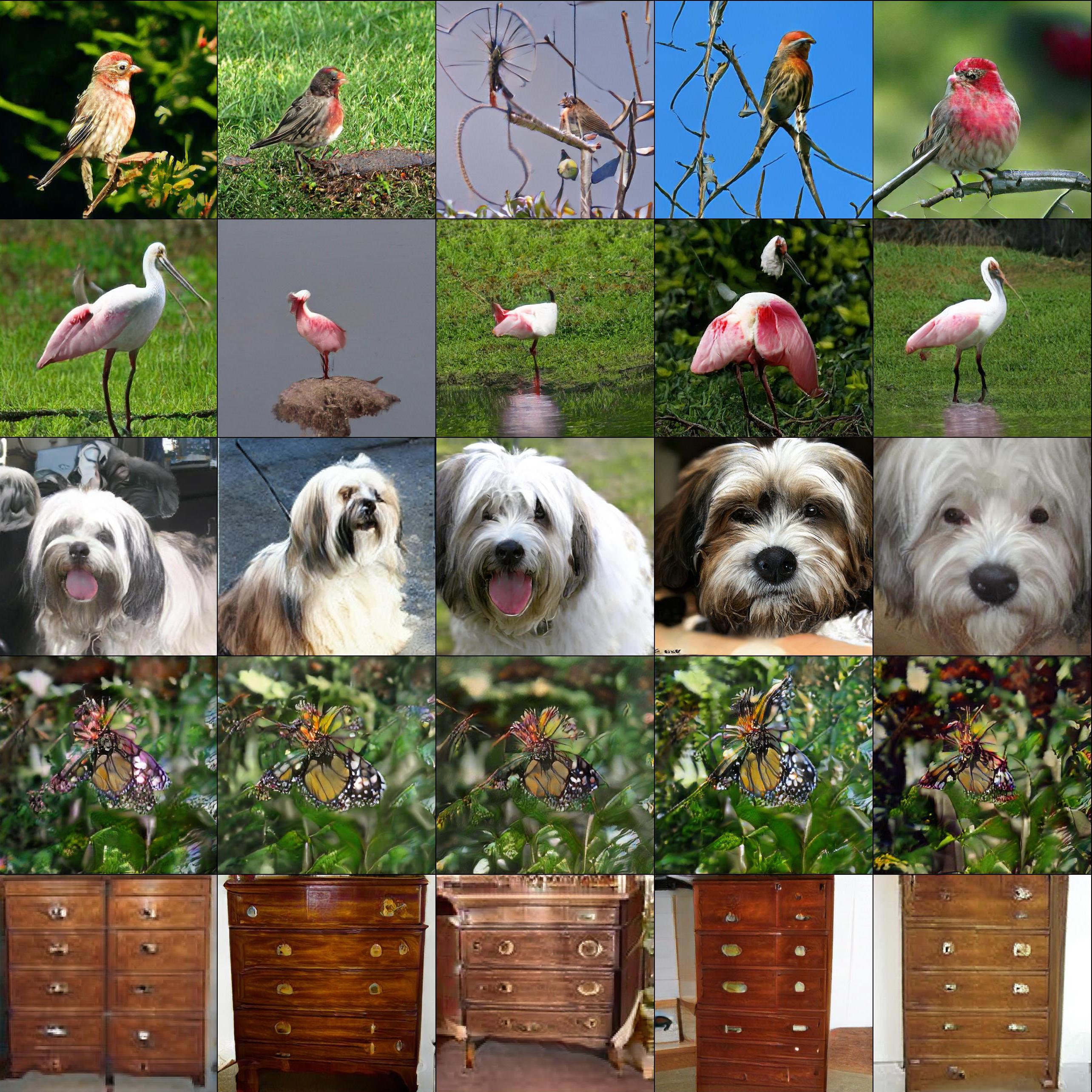}
\caption{\label{fig:biggan_512_samples_1}Class-conditional random samples at $512 \times 512$. Classes from the top row: 12 house finch, 129 spoonbill, 200 Tibetan terrier, 323 monarch butterfly, and 493 chiffonier.}
\end{figure}

\begin{figure}[t] 
\centering
\includegraphics[width=0.90\linewidth]{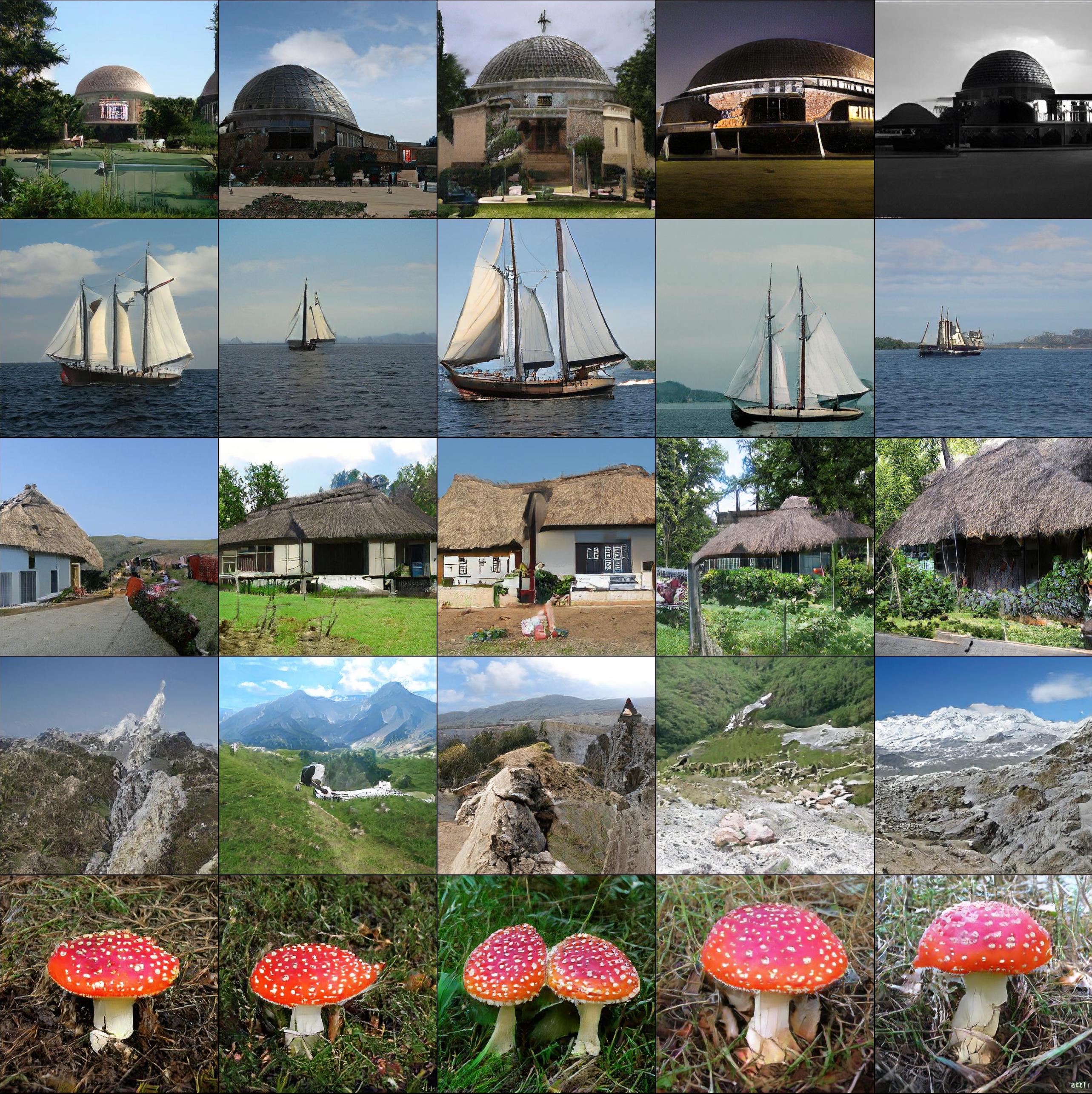}
\caption{\label{fig:biggan_512_samples_2}Additional class-conditional random samples at $512 \times 512$. Classes from the top row: 727 planetarium, 780 schooner, 853 thatch, 970 alp, and 992 agaric.}
\end{figure}

Refer to Figures \ref{fig:nsbgan_p_pretrained_samples_256} and \ref{fig:nsbgan_w_pretrained_samples_256} for class-conditional samples from \wtp and \wtw models at $256 \times 256$, respectively. Refer to Figures \ref{fig:nsbgan_p_pretrained_samples_512_1}, \ref{fig:nsbgan_p_pretrained_samples_512_2}, \ref{fig:nsbgan_w_pretrained_samples_512_1}, and \ref{fig:nsbgan_w_pretrained_samples_512_2} for class-conditional samples from \wtp and \wtw models at $512 \times 512$. Samples from BigGAN models at $256 \times 256$ and $512 \times 512$ are also included in \ref{fig:biggan_256_samples}, \ref{fig:biggan_512_samples_1}, and \ref{fig:biggan_512_samples_2} for comparison. We intentionally include random samples from the same classes for all the models to allow for direct and easy comparison.

\subsection{Full Resolution Samples from \wtw with Pre-trained Sampler}
\label{sec:wtw_full_res}
\begin{figure}[t] 
\centering
\includegraphics[]{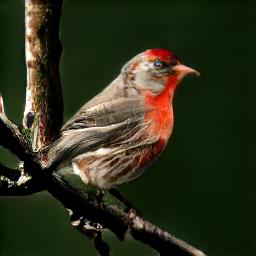}
\caption{\label{fig:nsbgan_w_sample_full256_1}Full $256 \times 256$ resolution sample from \wtw with pre-trained sampler.}
\end{figure}

\begin{figure}[t] 
\centering
\includegraphics[]{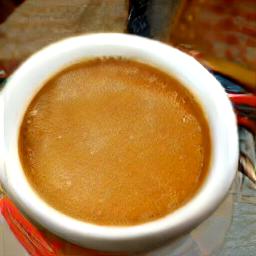}
\caption{\label{fig:nsbgan_w_sample_full256_2}Full $256 \times 256$ resolution sample from \wtw with pre-trained sampler.}
\end{figure}

\begin{figure}[t] 
\centering
\includegraphics[]{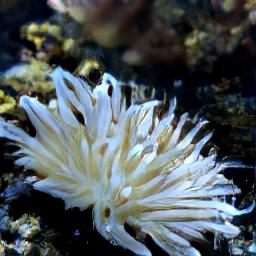}
\caption{\label{fig:nsbgan_w_sample_full256_3}Full $256 \times 256$ resolution sample from \wtw with pre-trained sampler.}
\end{figure}

\begin{figure}[t] 
\centering
\includegraphics[]{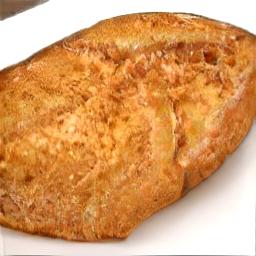}
\caption{\label{fig:nsbgan_w_sample_full256_4}Full $256 \times 256$ resolution sample from \wtw with pre-trained sampler.}
\end{figure}

Refer to Figures \ref{fig:nsbgan_w_sample_full256_1}, \ref{fig:nsbgan_w_sample_full256_2}, \ref{fig:nsbgan_w_sample_full256_3}, and \ref{fig:nsbgan_w_sample_full256_4} for full resolution samples from \wtw at $256 \times 256$. 

\begin{figure}[t] 
\centering
\includegraphics[width=1.0\linewidth]{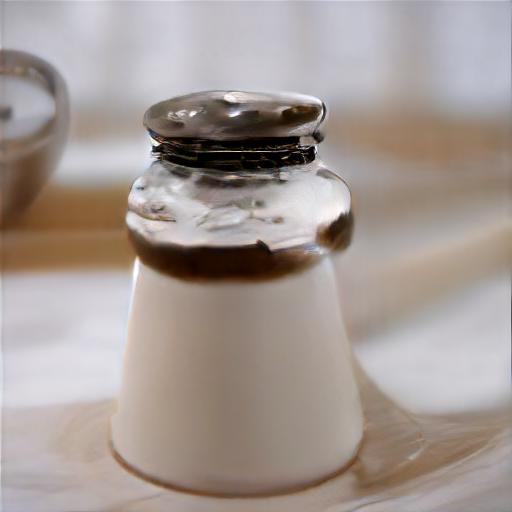}
\caption{\label{fig:nsbgan_w_sample_full512_1}Full $512 \times 512$ resolution sample from \wtw with pre-trained sampler.}
\end{figure}

\begin{figure}[t] 
\centering
\includegraphics[width=1.0\linewidth]{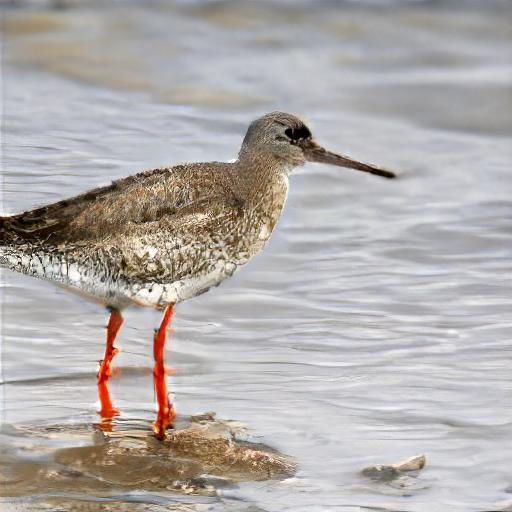}
\caption{\label{fig:nsbgan_w_sample_full512_2}Full $512 \times 512$ resolution sample from \wtw with pre-trained sampler.}
\end{figure}

\begin{figure}[t] 
\centering
\includegraphics[width=1.0\linewidth]{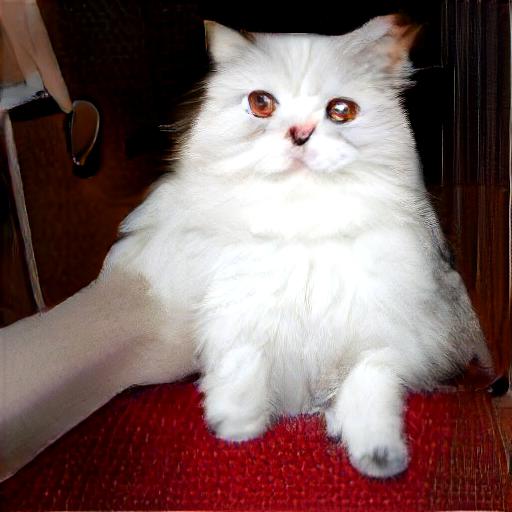}
\caption{\label{fig:nsbgan_w_sample_full512_3}Full $512 \times 512$ resolution sample from \wtw with pre-trained sampler.}
\end{figure}

\begin{figure}[t] 
\centering
\includegraphics[width=1.0\linewidth]{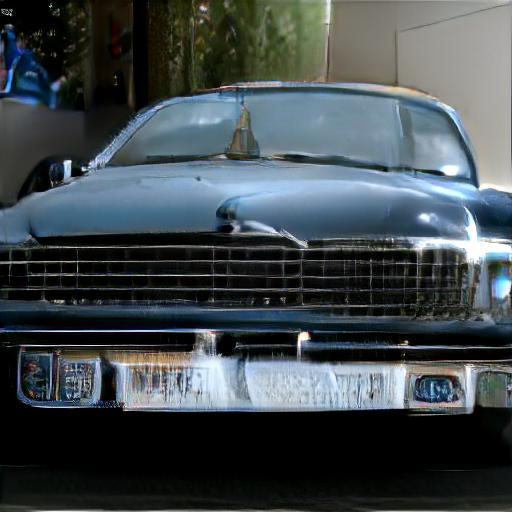}
\caption{\label{fig:nsbgan_w_sample_full512_4}Full $512 \times 512$ resolution sample from \wtw with pre-trained sampler.}
\end{figure}

Refer to Figures \ref{fig:nsbgan_w_sample_full512_1}, \ref{fig:nsbgan_w_sample_full512_2}, \ref{fig:nsbgan_w_sample_full512_3}, and \ref{fig:nsbgan_w_sample_full512_4} for $512 \times 512$. These samples are fitted to the page.

\subsection{Full Resolution Samples from \wtp with Pre-trained Sampler}
\label{sec:wtp_full_res}
\begin{figure}[t] 
\centering
\includegraphics[]{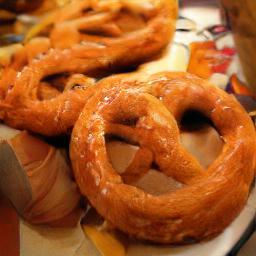}
\caption{\label{fig:nsbgan_p_sample_full256_1}Full $256 \times 256$ resolution sample from \wtp with pre-trained sampler.}
\end{figure}

\begin{figure}[t] 
\centering
\includegraphics[]{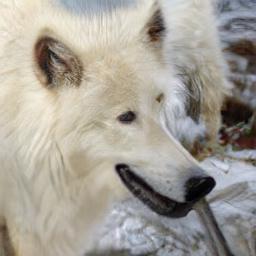}
\caption{\label{fig:nsbgan_p_sample_full256_2}Full $256 \times 256$ resolution sample from \wtp with pre-trained sampler.}
\end{figure}

\begin{figure}[t] 
\centering
\includegraphics[]{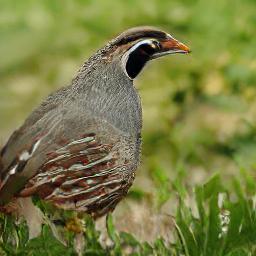}
\caption{\label{fig:nsbgan_p_sample_full256_3}Full $256 \times 256$ resolution sample from \wtp with pre-trained sampler.}
\end{figure}

\begin{figure}[t] 
\centering
\includegraphics[]{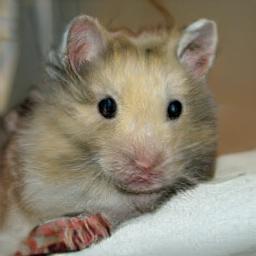}
\caption{\label{fig:nsbgan_p_sample_full256_4}Full $256 \times 256$ resolution sample from \wtp with pre-trained sampler.}
\end{figure}

Refer to Figures \ref{fig:nsbgan_p_sample_full256_1}, \ref{fig:nsbgan_p_sample_full256_2}, \ref{fig:nsbgan_p_sample_full256_3}, and \ref{fig:nsbgan_p_sample_full256_4} for full resolution samples from \wtw at $256 \times 256$. 

\begin{figure}[t] 
\centering
\includegraphics[width=1.0\linewidth]{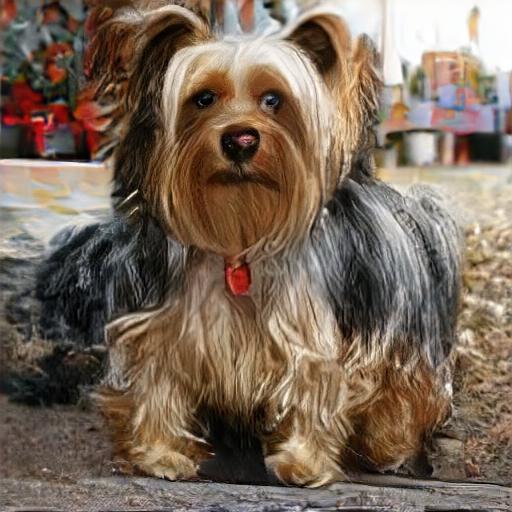}
\caption{\label{fig:nsbgan_p_sample_full512_1}Full $512 \times 512$ resolution sample from \wtp with pre-trained sampler.}
\end{figure}

\begin{figure}[t] 
\centering
\includegraphics[width=1.0\linewidth]{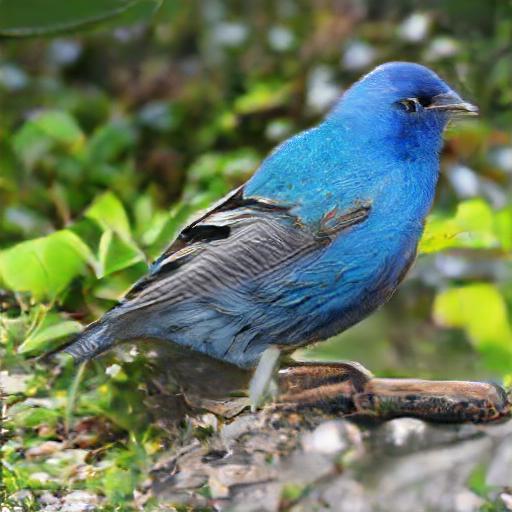}
\caption{\label{fig:nsbgan_p_sample_full512_2}Full $512 \times 512$ resolution sample from \wtp with pre-trained sampler.}
\end{figure}

\begin{figure}[t] 
\centering
\includegraphics[width=1.0\linewidth]{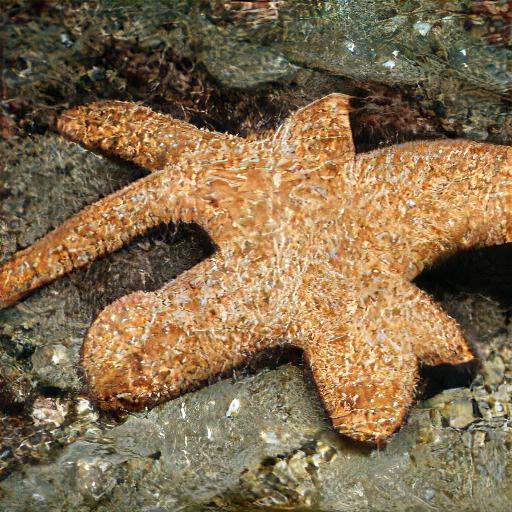}
\caption{\label{fig:nsbgan_p_sample_full512_3}Full $512 \times 512$ resolution sample from \wtp with pre-trained sampler.}
\end{figure}

\begin{figure}[t] 
\centering
\includegraphics[width=1.0\linewidth]{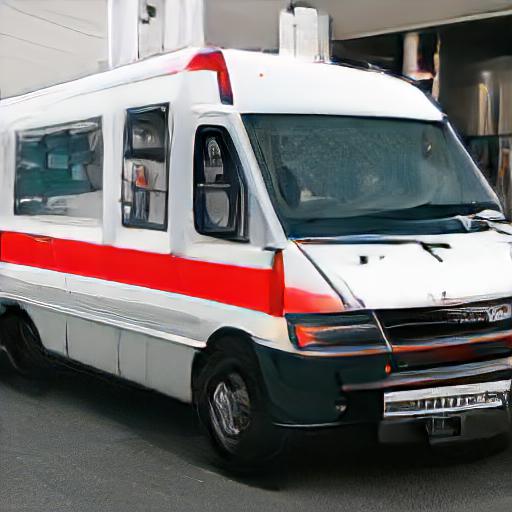}
\caption{\label{fig:nsbgan_p_sample_full512_4}Full $512 \times 512$ resolution sample from \wtp with pre-trained sampler.}
\end{figure}

Refer to Figures \ref{fig:nsbgan_p_sample_full512_1}, \ref{fig:nsbgan_p_sample_full512_2}, \ref{fig:nsbgan_p_sample_full512_3}, and \ref{fig:nsbgan_p_sample_full512_4} for $512 \times 512$. These samples are fitted to the page.

\section{Results with Truncation and Rejection Sampling}

\subsection{Learned Samplers}
\label{sec:learned_samplers_detailed}

For a more thorough analysis of our \wt models with learned samplers, we apply truncation and rejection sampling to study their effects on FID and IS. We apply truncation at $[1.0, 0.8, 0.6, 0.4, 0.2, 0.1]$, and, given the truncation level at which the model outputs the best FID, we apply rejection sampling with a pre-trained Inception V3 model on ImageNet. We test rejection sampling thresholds at 0.70, 0.80, 0.90, and 0.95. Refer to Table \ref{tab:learned_detailed} for quantitative results on FID and IS. We also include FID and IS results for UNet-W decoder that was only trained with a MSE loss.

\begin{table}[]
\centering
\begin{tabular}{@{}c|c|c|c|c@{}}
\toprule
 \textbf{Sampler} &  \textbf{Decoder}& \textbf{Resolution} & \textbf{min FID / IS} & \textbf{FID / max IS}\\ \midrule
Learned-P-64 & ESRGAN-P & 256 & 32.66 / 89.81 & 33.61 / 96.12\\ \midrule
Learned-W-64 & ESRGAN-W & 256 & 21.82 / 119.75 & 27.52 / 219.4\\ \midrule
Learned-P-64 & UNet-W & 256 & 35.60 / 84.68 &  38.58 / 129.4 \\ \bottomrule
\end{tabular}
\caption{\label{tab:learned_detailed} Minimum FID / IS (column 4) and FID / Minimum IS (column 5) attained with different levels of truncation [1.0, 0.8, 0.6, 0.4, 0.2, 0.1] and rejection sampling [0.70, 0.80, 0.90, 0.95].}
\end{table}

\subsection{Pre-trained Samplers}
\label{sec:pretrained_sampler_detailed}
We conduct the same detailed analysis for our \wt models with pre-trained samplers as the above. Refer to Table \ref{tab:pretrained_detailed} for quantitative results on FID and IS. 

\begin{table}[]
\centering
\begin{tabular}{@{}c|c|c|c|c@{}}
\toprule
 \textbf{Sampler} &  \textbf{Decoder}& \textbf{Resolution} & \textbf{min FID / IS} & \textbf{FID / max IS} \\ \midrule
Pretrained-128-64 & ESRGAN-P & 256 & 12.28 / 46.06 & 13.85 / 229.6 \\ \midrule
Pretrained-128-64 & ESRGAN-W & 256 & 12.66 / 45.54 & 20.44 / 285.5 \\ \midrule
Pretrained-256-128 & ESRGAN-P & 512 & 10.30 / 213.35 & 21.85 / 338.4 \\ \midrule
Pretrained-256-128 & ESRGAN-W & 512 & 10.59 / 52.14 & 22.26 / 332.7 \\ \bottomrule
\end{tabular}
\caption{\label{tab:pretrained_detailed} Minimum FID / IS (column 4) and FID / Minimum IS (column 5) attained with different levels of truncation [1.0, 0.8, 0.6, 0.4, 0.2, 0.1] and rejection sampling [0.70, 0.80, 0.90, 0.95].}
\end{table}

\section{Architecture, Hyperparameters, and Training Details}

\subsection{Sampler}
Refer to \ref{tab:arch_learned} for details about the architecture and hyperparameters of the learned samplers (Learned-P-64 and Learned-W-64). We attempted to train the sampler with both BigGAN and BigGAN-deep architectures and report on the models that achieved the best FID for each. Empirically, we found the models trained on the pixel-space much more unstable than on the wavelet-space, with four out of five models diverging.

\begin{table}[htb!]
\centering
\begin{tabular}{@{}ccc@{}}
\toprule
 & \textbf{Learned-W-64} & \textbf{Learned-P-64} \\ \midrule
\textbf{Best model type} & BigGAN-deep & BigGAN \\ \midrule
\textbf{Batch size} & 512 & 512 \\ \midrule
\textbf{Learning rate of generator} & 1e-4 & 1e-4 \\ \midrule
\textbf{Learning rate of discriminator} & 4e-4 & 4e-4 \\ \midrule
\textbf{Attention resolution} & 128 & 120 \\ \midrule
\textbf{Dimension of random noise ($z$ dim)} & 32 & 32 \\ \midrule
\makecell{\textbf{Number of  resblocks per stage in} \\  \textbf{generator/discriminator}}& 2 & 1 \\ \midrule
\textbf{Adam optimizer $\beta_{1}$} & 0 & 0 \\ \midrule
\textbf{Adam optimizer $\beta_{2}$} & 0.999 & 0.999 \\ \midrule
\textbf{Adam optimizer $\epsilon$} & 1e-8 & 1e-8 \\ \midrule
\textbf{Training iterations} & 250000 & 250000 \\ \bottomrule
\end{tabular}
\caption{\label{tab:arch_learned} Hyperparameters of learned samplers (wavelet and pixel)}
\end{table}

\subsection{\wt Decoders}
\label{sec:esrgan_decoders}

\paragraph{ESRGAN-W and ESRGAN-P}

A similar training procedure in ESRGAN is conducted for both of our decoders, ESRGAN-W and ESRGAN-P. First, SRResNet, with batch normalization removed, is trained with $L_{1}$ loss: $L_{1} = \mathbb{E}_{x_i}||G(x_{i})-y||_{1}$, where $x_{i}$ is the down-sampled image and $y$ is the target image.  After training with this $L_{1}$ loss for 150k iterations, the GAN is trained with perceptual and adversarial losses added. Therefore, the total loss for the generator becomes:
\begin{equation} \label{esrgan_loss}
L_{total} = L_{percep} + \lambda L_{G} + \eta L_{1}
\end{equation}

Perceptual loss is implemented with a pre-trained VGG-19 model on ImageNet and we use the features of the fourth layer before the fifth max-pooling layer. This GAN model is then trained for another 150k iterations.

As in the training for the learned sampler in wavelet domain, because the wavelet-encoded input image does not lie in the standard $[0, 1]$ range for ESRGAN-W, a normalization step is applied to transform the range into $[0, 1]$. By using the minimum and maximum values of the wavelet-encoded input images, pre-calculated in the pre-processing step, we use the following normalization technique: $x_{i_{norm}} = \frac{x_i + \ceil{|min_x|}}{\ceil{max_x}}$, where $x_i$ is the wavelet-encoded input and $min_x$ and $max_x$ are the pre-calculated minimum and maximum values of the wavelet-encoded input images.

Both ESRGAN-W and ESRGAN-P share the same architecture design, except for how the input is up-scaled and added to the learned features in order to induce the residual learning paradigm. Refer to Table \ref{tab:arch_esrgan_dec} for specific details of the architecture and hyperparameters.

\begin{table}[htb!]
\centering
\begin{tabular}{@{}cc@{}}
\toprule
 & \textbf{ESRGAN-W / ESRGAN-P} \\ \midrule
\textbf{Input size} & 64 $\times$ 64 \\ \midrule
\textbf{Batch size} & 32 \\ \midrule
\textbf{Learning rate} & 1e-4 \\ \midrule
\textbf{Number of residual blocks} & 16 \\ \midrule
\textbf{Conv filter size} & 3 \\ \midrule
\textbf{Adam optimizer $\beta_{1}$} & 0.9 \\ \midrule
\textbf{Adam optimizer $\beta_{2}$} & 0.99 \\ \midrule
\textbf{$L_1$ loss weight ($\eta$)} & 1e-2 \\ \midrule
\textbf{$L_G$ loss weight ($\lambda$)} & 5e-3 \\ \midrule
\textbf{Training iterations} & 150000 \\ \bottomrule
\end{tabular}
\caption{\label{tab:arch_esrgan_dec} Hyperparameters of ESRGAN decoders}
\end{table}

\subsection{UNet}
\label{sec:unet_decoder}

\paragraph{UNet in Wavelet Domain}
After a wavelet transform, the original image can be deterministically recovered from the TL, TR, BL, and BR patches using IWT. \wtw, however, discards the high-frequency patches (TR, BL, and BR) during its encoding step, rendering a fully deterministic decoding impossible. To resolve this, the \wt decoder first learns to recover the missing high-frequency patches (using a neural network) and then deterministically combines them using IWT to reconstruct the original input. 

Since \wt's encoding operation recursively applies wavelet transforms and discards the high-frequency components at each of the $L$ encoding levels, we train $L$ decoder networks to reconstruct the missing frequencies at each corresponding level of decoding. To parallelize the training of these decoder networks, we perform multiple wavelet transforms to the original high-resolution dataset, generating $L$ training sets of TL, TR, BL and BR patches. This allows us to independently train each of the decoder networks (in a supervised manner) to reconstruct the missing high-frequency patches conditioned on the corresponding TL.
This parallelization boosts the convergence rate of \wt, allowing for a fully trained decoder in under 48 hours.

Again with a slight abuse of notation, let $W_{1,1}^l$ be the TL patch at level $l$ and $\text{IWT} = \text{WT}^{-1}$. We can write the decoder for level $l$ as

\begin{align}
\label{eq:decode}
     D_{l}(W_{1,1}^l;\Theta) &= \text{IWT}\Bigg(
     \left[
        \begin{array}{c|c}
        W_{1,1}^l & f_{\theta_{1,2}}^l(W_{1,1}^l) \\
        \hline
        f_{\theta_{2,1}}^l(W_{1,1}^l) & f_{\theta_{2,2}}^l(W_{1,1}^l)
        \end{array}
     \right]
     \Bigg) \nonumber \\ 
     &\text{for }\quad 1 \leq l \leq L.
\end{align}
Here, each $f^l$ is a deep neural network that is trained to reconstruct one of the TR, BL, or BR patches at level $l$, conditioned on the TL patch ($W_{1,1}^l$). 

The main challenge in high-resolution image generation lies in overcoming the curse of dimensionality. But by leveraging IWT, the original dimensionality of the image is completely bypassed in the \wt decoder. Therefore, in practice, at each level, we further divide the TR, BL and BR patches by applying WT until we reach the patch dimensionality of 32x32. This can be done irrespective of the original dimensionality of the image. We then reconstruct these patches in parallel and recover the patches and the original image by recursively applying IWT. This ability of \wt to bypass the original dimensionality of the input separates it from all other SR methods that operate in the pixel space. 

\begin{figure}[t] 
\centering
\subfloat[\wt]{
    \includegraphics[width=0.50\linewidth]{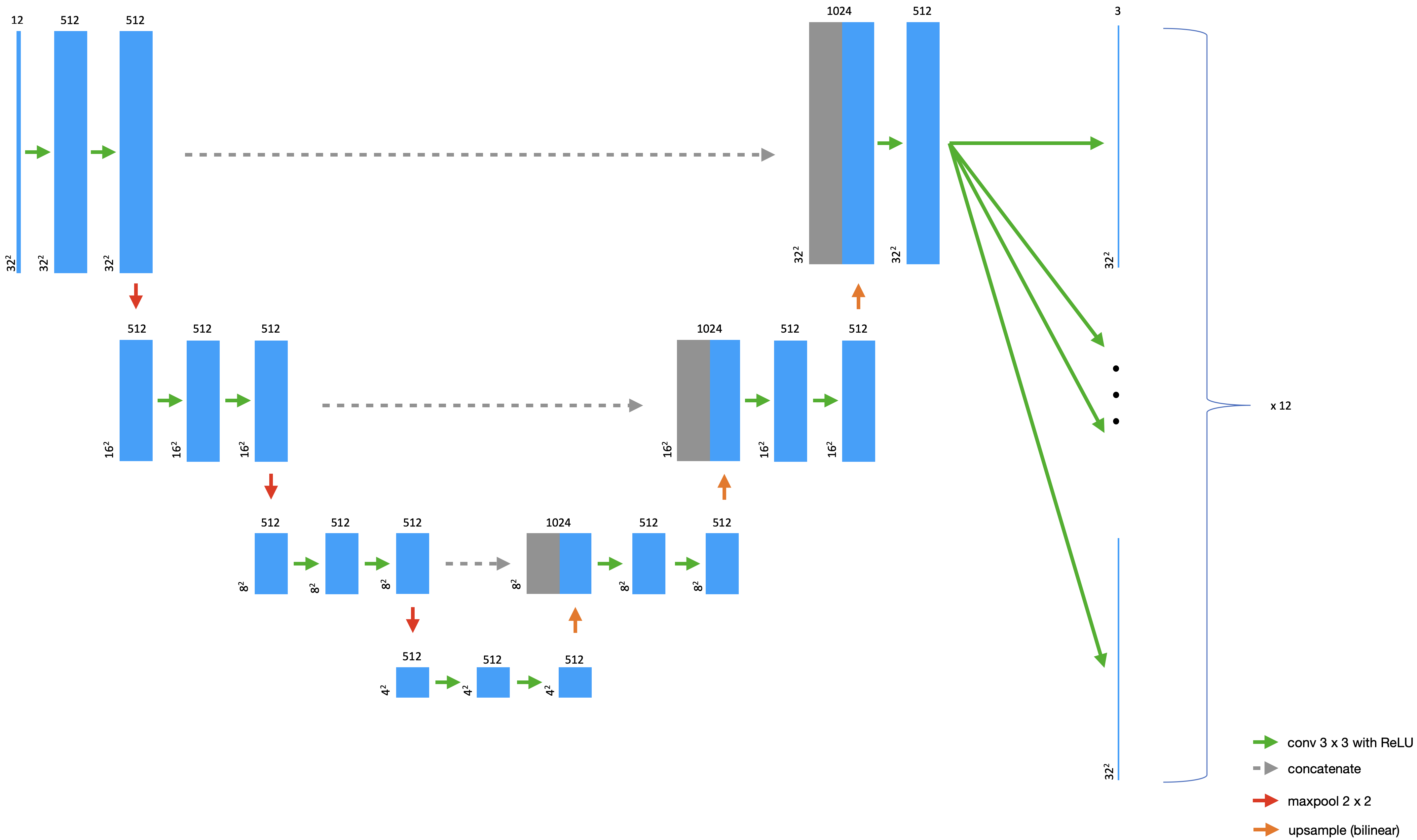}
    \label{fig:unet_wt}
}
\subfloat[\wtp]{
    \includegraphics[width=0.50\linewidth]{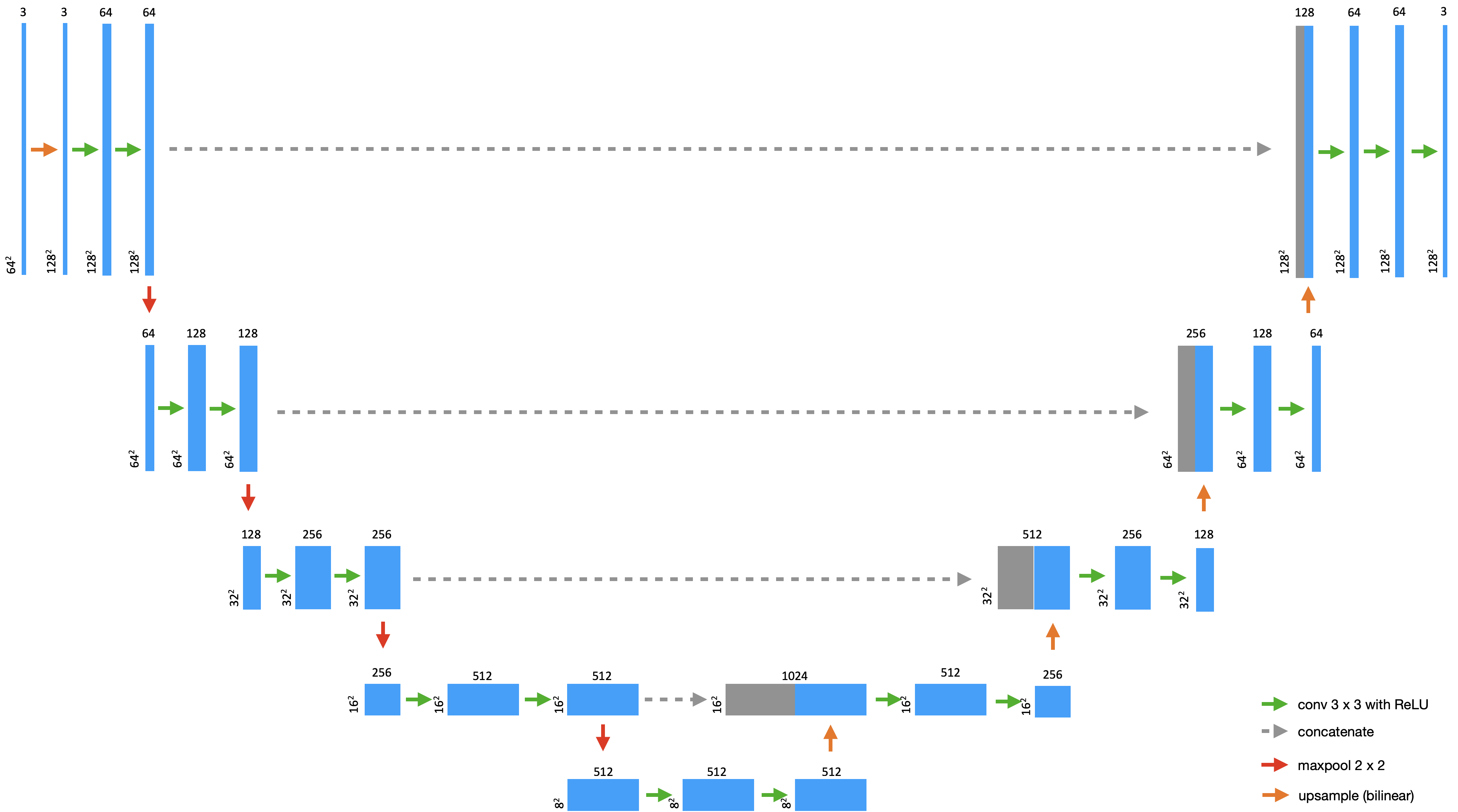}
    \label{fig:unet_pixel}
}
\caption{UNet-based decoding architecture for \wt and \wtp models. }
\label{fig:unet}
\end{figure}

Refer to Figure \ref{fig:unet_wt} and Table \ref{tab:arch_wt_dec} for architecture details and hyperparameters of the UNet decoder for the wavelet-space. 

\begin{figure}[t] 
\centering
\includegraphics[width=0.99\linewidth]{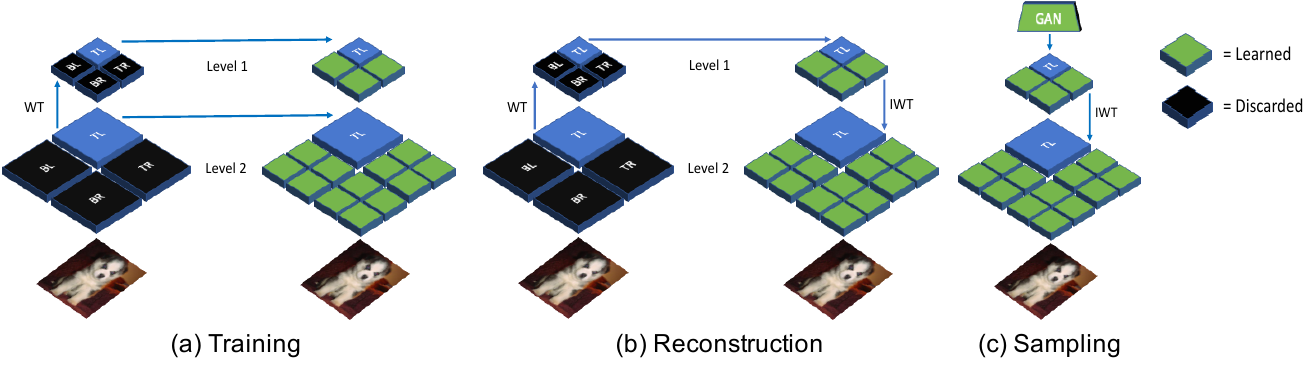}
\caption{\label{fig:scheme} \wt schematic for (a) training, (b) reconstruction and (c) sampling.}
\end{figure}

Refer to Figure \ref{fig:scheme} for the training schematic with the UNet decoders in the wavelet-space.

\paragraph{Architecture} We realize each of the $L$ decoder neural networks $f_{\theta_l}$ with a slightly modified version of the UNet architecture \citep{unet} rather than the commonly used transposed-convolution based architecture. UNet is typically used for image segmentation and object detection tasks. As shown in Figure \ref{fig:unet}, UNet is an autoencoder architecture that has skip-connections from each encoding layer to its corresponding decoding layer. These skip connections copy and paste the encoding layer's output into the decoding layer, allowing the decoder to exclusively focus on reconstructing only the missing, residual information. This architectural design makes it a compelling fit for decoding in \wt. We modify the UNet architecture by appending three shallow networks to its output with each one reconstructing one of the three high-frequency patches. This setup allows us to capture the dependencies between the high-frequency patches while also allowing sufficient capacity to capture to their individual differences.

\paragraph{UNet in Pixel Domain}

The UNet-P decoder is also a partly-learned function that uses a modified UNet-based architecture. First, the encoded image is deterministically upsampled using an interpolation-based method. This leads to a low-quality, blurry image at the same size as the original image. We then train a UNet to fill in the missing details in a similar approach to image super-resolution methods \citep{hu2019runet}. Unlike the \wt decoder that circumvents the CoD by avoiding reconstructions in the original data dimensionality, the \wtp decoder still has to operate on the full image size, resulting in a larger decoder.

Refer to Figure \ref{fig:unet_pixel} and Table \ref{tab:arch_pixel_dec} for architecture details and hyperparameters of the UNet decoder for the pixel-space.

\begin{table}[htb!]
\centering
\begin{tabular}{@{}ccc@{}}
\toprule
 & \textbf{Level 1 Decoder} & \textbf{Level 2 Decoder} \\ \midrule
\textbf{Input size} & 32 $\times$ 32 & 32 $\times$ 32 \\ \midrule
\textbf{Batch size} & 128 & 64 \\ \midrule
\textbf{Learning rate} & 1e-4 & 1e-4 \\ \midrule
\textbf{Layers} & 16 & 16 \\ \midrule
\textbf{Conv filter size} & 3 & 3 \\ \midrule
\textbf{Number of added shallow networks}& 12 & 48 \\ \midrule
\textbf{Adam optimizer $\beta_{1}$} & 0.9 & 0.9 \\ \midrule
\textbf{Adam optimizer $\beta_{2}$} & 0.999 & 0.999 \\ \midrule
\textbf{Adam optimizer $\epsilon$} & 1e-8 & 1e-8 \\ \midrule
\textbf{Training iterations} & 288000 & 232000 \\ \bottomrule
\end{tabular}
\caption{\label{tab:arch_wt_dec} Hyperparameters of UNet-W decoders (level 1 and level 2)}
\end{table}

\begin{table}[htb!]
\centering
\begin{tabular}{@{}ccc@{}}
\toprule
 & \textbf{Level 1 Decoder} & \textbf{Level 2 Decoder} \\ \midrule
\textbf{Input size} & 64 $\times$ 64 & 128 $\times$ 128 \\ \midrule
\textbf{Batch size} & 64 & 64 \\ \midrule
\textbf{Learning rate} & 1e-4 & 1e-4 \\ \midrule
\textbf{Layers} & 19 & 19 \\ \midrule
\textbf{Conv filter size} & 3 & 3 \\ \midrule
\textbf{Adam optimizer $\beta_{1}$} & 0.9 & 0.9 \\ \midrule
\textbf{Adam optimizer $\beta_{2}$} & 0.999 & 0.999 \\ \midrule
\textbf{Adam optimizer $\epsilon$} & 1e-8 & 1e-8 \\ \midrule
\textbf{Training iterations} & 615000 & 178000 \\ \bottomrule
\end{tabular}
\caption{\label{tab:arch_pixel_dec} Hyperparameters of UNet-P decoders (level 1 and level 2)}
\end{table}

\section{\wt vs \wtp Information Content}
\label{sec:wtvswtp}
As we show in Figure \ref{fig:interpol_comp}, the information content in the latent embedding of \wt and \wtp are drastically different. The TL patches from \wt preserve more structural information in the image than the down-sampled image from \wtp. It is evident in the figure that pixel-based down-sampling method misses key structures and features, such as the face of the person, structure of the boat, and structure of the accordion keyboard, whereas wavelet encoding does not.

\section{Evaluation on LSUN Church with StyleGAN-2}
\label{sec:stylegan}
To demonstrate that the \wt paradigm can be different samplers and that our decoder trained on ImageNet generalizes fairly well to an unseen dataset, we conduct a set of experiments with the BigGAN sampler replaced with a StyleGAN-2 sampler. We test the StyleGAN-2 sampler with two decoders: one trained on ImageNet and another trained on LSUN Church. Illustrated in \ref{tab:style}, as expected, the decoder that was only trained on the LSUN Church dataset outperforms our decoder trained on ImageNet. However, even with our decoder, we reach a very competitive FID of 13.53, beating other recent two-step approaches, such as \citep{acceleration}.

\begin{table}[]
\centering
\begin{tabular}{@{}c|c|c|c|c@{}}
\toprule
 \textbf{Sampler} &  \textbf{Decoder}& \textbf{Resolution} & \textbf{FID} & \textbf{IS} \\ \midrule
\textbf{Pretrained-256-64} & ESRGAN-W & 256 & 13.53
 & 3.182 \\ \midrule
\textbf{Pretrained-256-64} & ESRGAN-W (Church) & 256 & 7.886 & 2.784 \\ \bottomrule
\end{tabular}
\caption{\label{tab:style} The above pre-trained samplers are StyleGAN-2 model trained on LSUN Church dataset. Even our decoder trained only on ImageNet reaches a competitive FID of 13.53, but evidently can be improved by training a new decoder on LSUN Church dataset. These results show that our \wt models can work with different samplers than BigGAN and that our decoder generalizes fairly well to unseen dataset. Therefore, we do not go about proving the compute efficiency in this case.}
\end{table}

\section{Slicing}
\label{sec:sl}
\begin{figure}[htb!] 
\centering
\subfloat[SPN Slicing Operation]{
    \includegraphics[width=0.45\linewidth]{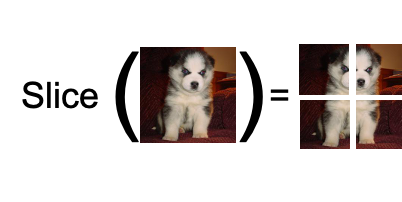}
    \label{fig:spn_s}
}
\subfloat[\wt Slicing Operation]{
    \includegraphics[width=0.45\linewidth]{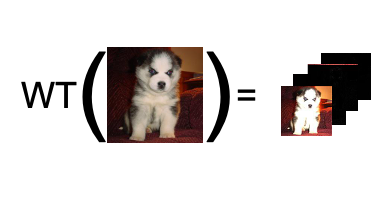}
    \label{fig:wt_s}
}
\caption{Unlike SPN, \wt slices the images in the frequency domain. As a result each patch contains the entire global structure of the input image. This helps alleviate any long-term dependency issues. }
\label{fig:slice}
\end{figure}

Figure \ref{fig:wt_s}, shows the difference between the slicing operations of SPN and \wt. SPN slices the images across the spatial dimensions thus introducing long-term dependencies. In contrast, \wt decomposes the image along different frequency bands that preserves the global structure of the image in each of the patches.





\section{Wavelet Transform.}
Wavelet transformation of an image (illustrated in Figure \ref{fig:wt}) is a two-step recursive process that splits the original image into four equal-sized patches, each a representation of the entire image in different frequency bands.  In the first step, a low-pass filter (LPF) and a high pass-filter (HPF) are applied to the original image. This produces two patches of the same size as the original image. Since the application of LPF and HPF leads to redundancy, we can apply Shannon-Nyquist theorem to downsample these patches by half without losing any information. In step two, the same process is repeated on the output of step one, splitting the original image into four equally-sized patches (TL, TR, BL and BR). TR, BL and BR contain increasingly higher frequencies of the input image, preserving horizontal, vertical and diagonal edge information, respectively (contributing to the sharpness of the image).

\section{\wt at Higher Resolution}
\label{sec:imagenet512}
\wt beats the BigGAN baseline model on FID at 512$\times$512 resolution. We hypothesize that this difference in performance is due to the fact that ImageNet dataset at 512 $\times$ 512 is generated by using pixel-based interpolation up-sampling of the original data. Since BigGAN is trained to generate this blurry data, compared to our \wt approach, which uses a pre-trained sampler at 256 $\times$ 256 and uses a learned super-resolution model to up-sample to 512 $\times$ 512, it performs sub-optimally. 

\end{document}